\documentclass[twocolumn]{aastex631}

\submitjournal{ApJ}

\shorttitle{X-ray GRB}
\shortauthors{Chen et al.}

\usepackage{amsmath}
\usepackage{float}
\usepackage{tabularx}
\usepackage{multirow}
\usepackage{xcolor}

\begin{document}

\title{X-ray Emission Signatures of Neutron Star Mergers}

\author[0000-0001-.7880-2407]{Connery J. Chen}
\affiliation{Nevada Center for Astrophysics, University of Nevada, 4505 S. Maryland Pkwy., Las Vegas, NV 89154-4002, USA}
\affiliation{Department of Physics and Astronomy, University of Nevada, 4505 S. Maryland Pkwy., Las Vegas, NV 89154-4002, USA}

\author[0000-0002-8614-8721]{Yihan Wang}
\affiliation{Nevada Center for Astrophysics, University of Nevada, 4505 S. Maryland Pkwy., Las Vegas, NV 89154-4002, USA}
\affiliation{Department of Physics and Astronomy, University of Nevada, 4505 S. Maryland Pkwy., Las Vegas, NV 89154-4002, USA}

\author[0000-0002-9725-2524]{Bing Zhang}
\affiliation{Nevada Center for Astrophysics, University of Nevada, 4505 S. Maryland Pkwy., Las Vegas, NV 89154-4002, USA}
\affiliation{Department of Physics and Astronomy, University of Nevada, 4505 S. Maryland Pkwy., Las Vegas, NV 89154-4002, USA}

\begin{abstract}
Neutron star (NS) mergers, including both binary NS mergers and black hole-NS mergers, are multimessenger sources detectable in both gravitational wave (GW) and electromagnetic (EM) radiation. The expected EM emission signatures depend on the source's progenitor, merger remnant, and observer's line of sight (LoS). Widely discussed EM counterparts of NS mergers have been focused in the gamma-ray (in terms of short-duration gamma-ray bursts) and optical (in terms of kilonova) bands. 
In this paper, we demonstrate that X-ray emission provides a powerful and complementary probe of post-merger physics and geometry, offering diagnostic signatures across both the prompt and long-term afterglow phases.
We consider several binary progenitor and central engine models and investigate X-ray emission signatures from the prompt phase immediately after the merger to the afterglow phase extending years later.
For the prompt phase, we devise a general method for computing phenomenological X-ray light curves and spectra for structured jets viewed from any LoS, which can be applied to X-ray observations of NS mergers to constrain the geometry.
The geometric constraints can in turn be used to model the afterglow and estimate a peak time and flux---to preemptively determine afterglow characteristics would be monumental for follow-up observation campaigns of future GW sources.  
Finally, we provide constraints on the time window for X-ray counterpart searches of NS mergers across a range of luminosity distances and detector sensitivities.
\end{abstract}

\keywords{gamma-ray burst: general -- stars: magnetars -- methods: numerical -- X-rays: stars}

\section{Introduction}
\label{sec:intro}

The first multimessenger detection of a binary neutron star (BNS) merger in 2017 through both GW \citep{GW170817} and electromagnetic (EM) observations \citep{170817, grb170817_fermi, grb170817_integral, troja+17_GW170817_xray, valenti_170817_kn, covino_170817_kn, hallinan_17,zhang+18_peculiar_grb} opened a new chapter in observational multimessenger astrophysics. The source continues to exhibit remarkable X-ray properties years after the merger, enabling stringent tests of physical models~\citep{troja+20_170817, hajela_22}.

In general, a neutron star (NS) merger is defined as consisting of at least one NS in the merger system. Such systems, including both NS-NS mergers (also called binary NS mergers, or BNS mergers), and black hole (BH)-NS mergers, are ideal multimessenger systems detectable through both GW and EM signals. Leading up to an NS merger, the binary system loses orbital energy through GW radiation, causing the binary separation to decrease. The final outcome and subsequent GW and EM signals depend on the initial parameters of the system and the NS EoS, which is poorly constrained~\citep[for recent reviews, see e.g.,][]{lattimer21_ns_eos, burgio+21_ns_eos}. 
For BNS mergers, the radioactive decay of r-process elements synthesized in the neutron-rich material ejected during the merger can power a luminous EM transient which peaks in the optical/infrared band, termed a ``kilonova''~\citep{li_98, kulkarni_05, metzger+10_kilonova, kasen_13}. 
In the case of BH-NS mergers, if the mass ratio between the BH and NS is not too large, and/or the BH is spinning rapidly, and/or the NS EoS is stiff, then the NS can be tidally disrupted outside the BH innermost stable circular orbit, possibly launching jets and leaving behind neutron-rich ejecta that may power a kilonova. Otherwise, the NS will plunge directly into the BH without significant mass ejection, resulting in little to no EM counterparts. For both BNS mergers and BH-NS mergers with tidal disruption, relativistic jets can be launched perpendicular to the orbital plane, which may be detected as a gamma-ray burst (GRB). 

GRBs are the most luminous transients in the universe and are produced during violent explosions and cataclysmic events that launch highly relativistic jets \citep[see e.g.][]{zhang18_grb_book}. Phenomenologically, they are classified into two types based on duration~\citep{kouveliotou_93}: short GRBs---with durations less than 2 seconds---and long GRBs---with durations greater than 2 seconds. Multi-wavelength and multimessenger observations have established that long-duration GRBs are typically produced during the core-collapse of massive stars, as supported by their association with Type Ic supernovae~\citep{woosley_bloom_06_grb_sne}. In contrast, short-duration GRBs are typically associated with NS mergers, as demonstrated by the coincident detection of a GW event from a BNS merger (GW170817) and a GRB (GRB170817A)~\citep{170817}\footnote{
Unlike typical short GRBs, EM observations of GRB170817A~\citep{grb170817_fermi, grb170817_integral,zhang+18_peculiar_grb} revealed that it possessed unusually low luminosity and soft spectral characteristics. Numerical models have suggested these observations to be consistent with a narrow jet viewed off-axis~\citep{mooley_18, troja_19, ghirlanda_19_170817, ryan_24}, or significant cocoon emission~\citep{gottlieb_18, sadeh_waxman_24}. This work studies the former scenario and obtains consistent results.}.
Numerous ``peculiar'' GRBs---long-duration GRBs with no associated SNe~\citep{dellavalle_06, fynbo_06, gehrels_06, galyam_06, yang_15_060614} and short-duration GRBs likely with massive star origins~\citep{antonelli_09, levesque_10, xin_11, rossi_22}---led \cite{zhang_06_type} and \cite{zhang_07_type} to propose that physical classifications do not always align with phenomenological ones, and that one should define GRBs based on their physical origins, i.e., Type I GRBs that originate from compact-object mergers, and Type II GRBs that are associated with the core-collapse of massive stars. The need of such a classification scheme is recently strengthened with the detection of two additional long-duration GRBs associated with kilonovae \citep{rastinejad_22, yang+22_peculiar_grb, levan_24, yang_24, sun_25} and one short-duration GRB associated with a supernova \citep{ahumada_21,zhangbb_21,rossi_22}. As noted by \cite{zhang_25}, GRB duration is shaped not only by the progenitor system, but also by properties of the central engine, the emission site, and the system's geometry. In this paper, we adopt the generalized classification based on progenitor systems and label GRBs associated with NS mergers ``Type I GRBs''.

Traditionally, the EM counterparts of NS mergers are believed to be most efficiently searched in the gamma-ray (for GRB emission) and optical (for kilonova emission) bands~\citep{metzger_berger_12}. Broadband afterglow emissions for both the GRB and kilonova are also expected. \cite{zhang_13} proposed that GW sources due to BNS mergers can be followed by bright long-duration X-ray emission if the merger product is a long-lived magnetar instead of black hole. Such an X-ray counterpart can be detected in the absence of a short GRB if the line of sight (LoS) is far from the jet axis (see also \cite{sun_17, sun_19}). The postulated magnetar engine would enhance the brightness of the kilonova \citep{yu_13} and the broadband afterglow \citep{gao_13}. Candidate GRB-less X-ray transients potentially originating from binary neutron star (BNS) mergers have been reported \citep[e.g.,][]{xue_19,bauer_17,quirola-vasquez_24,ai-zhang_21}, along with kilonova candidates possibly boosted by a magnetar engine \citep[e.g.,][]{gao_15,gao_17}. Advancing multimessenger observations of future binary neutron star merger events is vital for probing their merger remnants and observational geometries, and relies on advancements in sophisticated theoretical modeling.

Motivated by recent advances in X-ray detection capabilities---especially those enabled by fast X-ray transient (FXT) searches with the Einstein Probe \citep{EP} and SVOM \citep{svom}---in this work we focus on the X-ray band and investigate how X-ray emission signatures vary with binary progenitors, central engine remnants, and observer viewing angles. We develop a general method to model prompt X-ray emission from various NS mergers, and release the associated open-source code \texttt{PromptX}~\citep{promptx}. We present general prompt X-ray light curves and spectra that can be used to guide follow-up observations. We also employ the high-performance GRB afterglow modeling toolkit \texttt{VegasAfterglow}~\citep{wang_25_vegasafterglow} in generating model X-ray afterglows and investigate their consistency with prompt emission. The goal of this work is a comprehensive framework for modeling X-ray emissions from NS mergers across observational epochs.

In Section~\ref{sec:engine} we review the evolution of NS mergers starting from various progenitors, explore possible merger remnants, consider theoretical models for long-lived central engines, and examine how viewing geometry affects observational results.
In Section~\ref{sec:emissions} we outline observational characteristics of emission signatures relevant to NS mergers, including prompt, wind, and afterglow emissions. We introduce \texttt{PromptX}~\citep{promptx}, an open-source code to model prompt X-ray emission from Type I GRBs and a spindown-powered X-ray wind (for the case of a long-lived magnetar central engine). We calculate X-ray signatures for various LoS and produce corresponding X-ray afterglow light curves. 
In Section~\ref{sec:results} we present observational expectations of NS mergers from typical progenitors and their plausible central engine remnants viewed from any angle. We provide a summarized workflow to outline how one can seamlessly adopt these results to observational campaigns in X-rays.
We conclude our findings in Section~\ref{sec:conclusion}.

\section{Central Engine}\label{sec:engine}

\begin{figure}
    \centering
    \includegraphics[width=\linewidth]{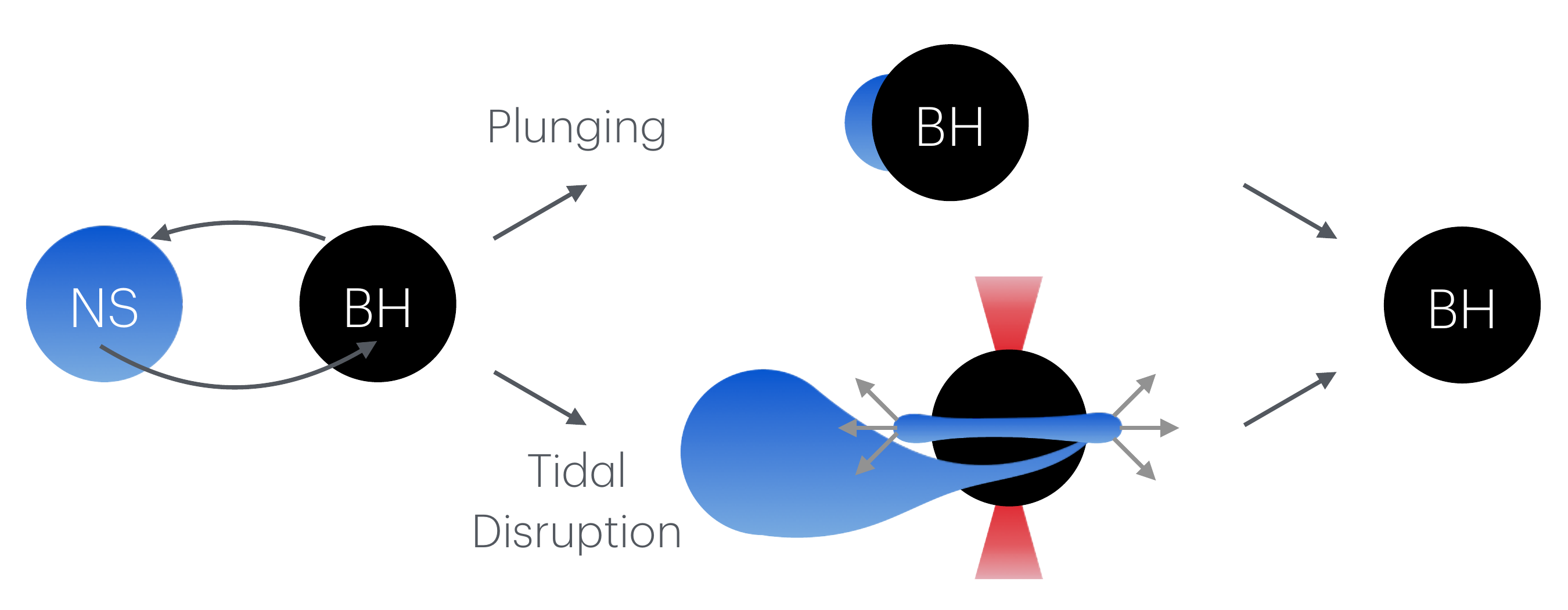}\\
     \vspace{1cm} 
    \includegraphics[width=\linewidth]{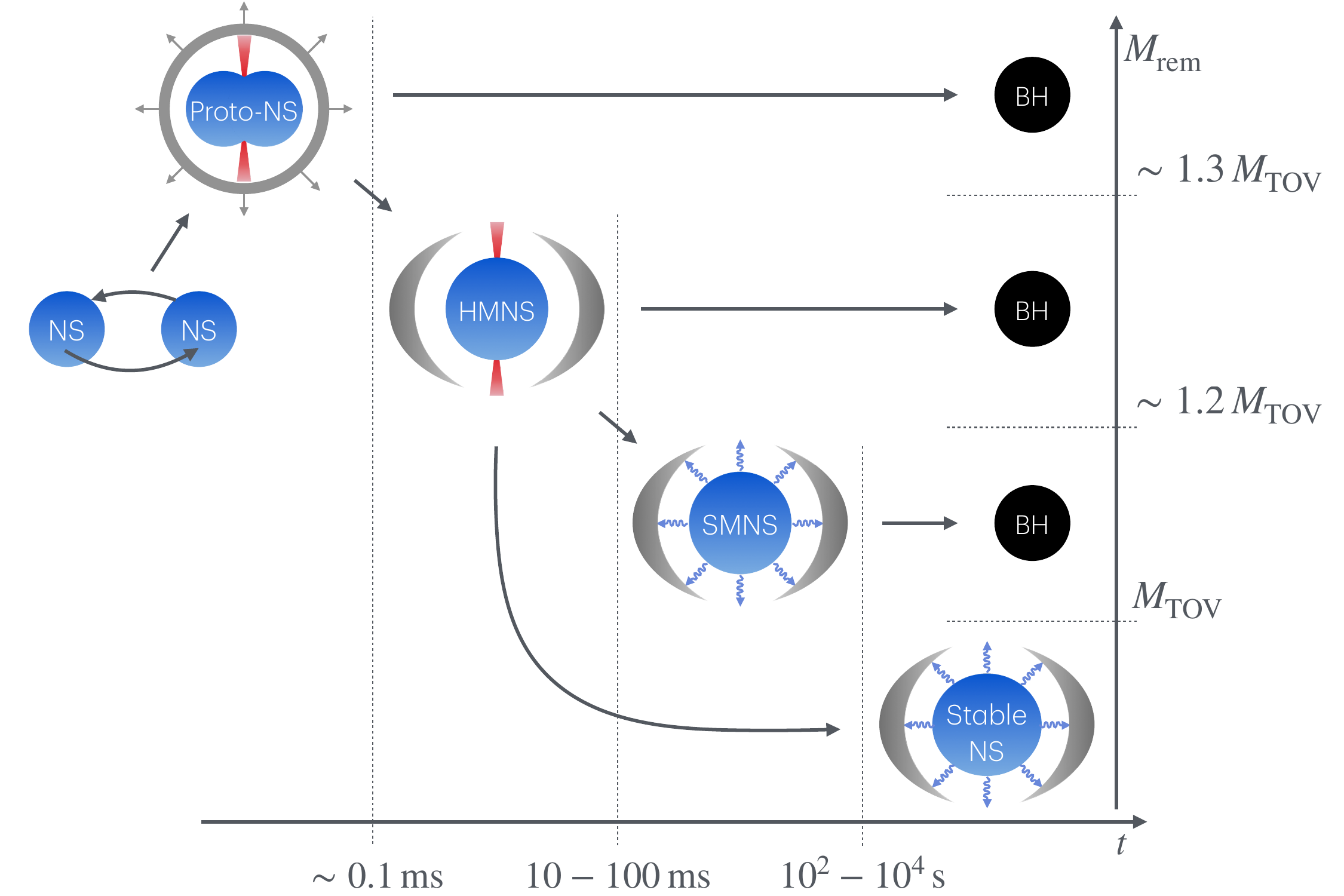}
    \caption{Central engine formation from a BH-NS merger (\textit{upper panel}) and a BNS merger (\textit{lower panel}).}
    \label{fig:central-engine}
\end{figure}

In compact object mergers, the accompanying high-energy EM emission is typically interpreted as arising from relativistic outflows powered by a central engine, such as a hyper-accreting black hole or a newly formed, rapidly rotating, highly magnetized neutron star (a magnetar).
Observations of the gamma-ray signal associated with GW170817 are consistent with this general picture. Alternative interpretations, such as emission arising from the breakout of a broad, mildly relativistic cocoon \citep{kasliwal_17, gottlieb_18}, remain possible.

The BH-NS merger product is always a black hole; thus, the formation of a hyper-accreting black hole central engine depends on whether the neutron star undergoes tidal disruption before plunging into the black hole (Fig.~\ref{fig:central-engine}, upper panel). The BNS merger product can either be a BH or long-lived magnetar central engine, depending on the total remnant mass and equation of state (Fig.~\ref{fig:central-engine}, lower panel; see also \citealt{gao+16_binary_ns, margalit_19, ai_20_eos}).

\subsection{Black hole}\label{sec:prompt_bh}
For BH-NS mergers, if certain conditions are met (e.g., comparable masses, eccentricity, stiff NS EoS, etc.), the NS can be tidally disrupted. 
In this scenario, part of the disrupted stellar material falls back to form an accretion disk around the black hole, which can power relativistic jets~\citep{janka_99_grb, rezzolla_11_grb, liu17_ndaf, ruiz_21, gottlieb_23_nsbh_grb, gottlieb_23_bhns, zhu_22_nsbh_grb} via the Blandford-Znajek mechanism~\citep{bz_77} or neutrino-antineutrino annihilation~\citep{zalamea_beloborodov_11} and produce late-time X-ray flares~\citep{Perna_06_flares, proga_zhang06_flares}; the remaining material is ejected, potentially contributing to kilonova emission \citep{huang_18, zhu_20, gompertz_23, kawaguchi_24}.

For a non-spinning BH, the approximate criterion for tidal disruption of the NS is given by
\begin{eqnarray}
r_{\rm TD}\sim\bigg(\frac{3M_\bullet}{M_*}\bigg)^{1/3}R> r_{\rm ISCO} \sim \frac{6GM_\bullet}{c^2},
\end{eqnarray}
where $r_{\rm TD}$ is the tidal disruption radius of the NS, $r_{\rm ISCO}$ is the innermost stable circular orbit of the black hole, $M_\bullet$ and $M_*$ are the masses of the BH and NS, respectively, and $R$ is the radius of the NS. 
For a rotating (Kerr) black hole with dimensionless spin parameter $a_*$, the ISCO radius depends on the spin and whether the orbit is prograde or retrograde:
\begin{eqnarray}
r_{\rm ISCO}(a_*) = \frac{GM_\bullet}{c^2} \Big[ 3 + Z_2 \mp \sqrt{(3 - Z_1)(3 + Z_1 + 2Z_2)} \Big],
\end{eqnarray}
where the upper (minus) sign corresponds to prograde orbits ($a_* > 0$), the lower (plus) sign to retrograde orbits ($a_* < 0$), and
\begin{eqnarray}
Z_1 &=& 1 + (1 - a_*^2)^{1/3} \Big[ (1 + a_*)^{1/3} + (1 - a_*)^{1/3} \Big],\\
Z_2 &=& \sqrt{3 a_*^2 + Z_1^2}.
\end{eqnarray}
In the maximal spin cases, $r_{\rm ISCO}$ ranges from $GM_\bullet/c^2$ for a prograde orbit ($a_*=+1$) to $9GM_\bullet/c^2$ for a retrograde orbit ($a_*=-1$). Therefore, tidal disruption of the NS is more feasible in a prograde system with a rapidly-rotating BH.

For BNS mergers, immediately before merger, some material with total mass $M_{\rm ej}$ will be dynamically ejected from the system, and a remnant with total mass $M_{\rm rem}$ will remain. If $M_{\rm rem} \gtrsim 1.3 \, M_{\rm TOV}$~\citep{ai_20_eos, metzger+10_kilonova}, a BH will be promptly formed within the dynamical timescale,
\begin{equation}
    t_{\rm dyn} = \left(\frac{R^3}{GM} \right)^{1/2} \approx 0.06 \left(\frac{R}{10\text{km}}\right)^{3/2}\left(\frac{M}{2M_\odot}\right)^{-1/2} \rm ms.
\end{equation}

\subsection{Magnetar}
While many models propose a hyper-accreting black hole as the central engine responsible for powering a GRB, \cite{duncan_thompson_92_magnetar} and \cite{usov_92} were among the first to propose that a newly formed, rapidly spinning, highly magnetized neutron star---a magnetar---could power a GRB by its enormous magnetic and spin energy reservoirs. In the context of BNS mergers, \cite{dai_06} and \cite{gao-fan_06} proposed that a supramassive, differentially rotating millisecond magnetar remnant could power extended X-ray activity in some Type I GRBs, contingent on a sufficiently stiff NS EoS. 
In regards to the GRB prompt emission itself, however, early theoretical and numerical studies have demonstrated that baryon loading in the polar regions severely limits the formation of clean, ultra-relativistic jets~\citep{mb_14, ciolfi_17}. However, more recent GRMHD simulations indicate that hypermassive neutron stars (discussed further in Section~\ref{sec:hmns}) can drive mildly relativistic, magnetically dominated polar outflows, which may be further accelerated to larger Lorentz factors compatible with GRBs, suggesting a potential pathway to overcome baryon-loading limitations~\citep{mosta_20, most23_jet, musolino_25}.

Notably, some Type I GRBs are followed by softer extended emission~\citep{norris_06}, and a significant fraction of Type I GRBs exhibit ``internal'' X-ray plateaus---a distinct phase characterized by shallow to flat decay followed by a very steep temporal decay which cannot be interpreted within the framework of external shock models~\citep{rowlinson_10, rowlinson_13}. These features can instead be naturally explained by continued energy injection from the spindown of a magnetar engine, with the sudden decay signaling its collapse to a BH~\citep{troja_07, metzger_08,zhang_14,lu15_ms_magnetar, siegel_ciolfi_16_1, siegel_ciolfi_16_2, sun_25}. Interestingly, kilonova-associated GRBs appear to populate a systematically dimmer region of the X-ray plateau luminosity--duration plane~\citep{dainotti+20}.
Some theoretical studies have explored possible explanations for X-ray plateaus in short GRBs that do not involve a magnetar engine~\citep{kisaka_ioka_15, lu_quataert_23,hamidani_24, lenart_25} or that consider purely geometric effects~\citep{oganesyan_20, beniamini_15}.
However, to account for the full Type I GRB dataset, a magnetar engine is apparently required for at least for some GRBs. 

If a BNS merger remnant is not too massive to promptly collapse into a BH ($M_{\rm rem} \lesssim 1.3 M_{\rm TOV}$), it may evolve through transient phases (Fig.~\ref{fig:central-engine}, lower panel). 

\subsubsection{Hypermassive Neutron Stars} \label{sec:hmns}

If $M_{\rm rem} \sim (1.2-1.3) \, M_{\rm TOV}$, the remnant can withstand gravitational collapse by rapid differential rotation---the defining characteristic of a hypermassive neutron star (HMNS). 
Differential rotation is not sustainable indefinitely---as the NS spins, magnetic field lines are wound into strong toroidal components, launching Alfv\'en waves that redistribute angular momentum outward and reduce the degree of differential rotation. This process acts on the Alfv\'en timescale~\citep{shapiro00},
\begin{equation}
\label{eq:t_alfven}
    t_A = \frac{R}{v_A} \\ \approx 1 \left(\frac{B_0}{10^{14} \text{ G}}\right)^{-1} \left(\frac{R}{10 \text{ km}}\right)^{-1/2} \left(\frac{M}{2 M_\odot}\right)^{1/2} \text{ s},
\end{equation}
where $v_A = B/(4\pi\rho^2)^{1/2}$ is the Alfv\'en speed. 
As magnetic braking reduces the degree of differential rotation, the HMNS loses centrifugal support. If the remnant mass still exceeds the maximum mass for a uniformly rotating NS (the supramassive limit), the remnant will promptly collapse to a black hole within the timescale of $t_A$.

The idea that a Type I GRB jet may be launched during or shortly after the HMNS phase has been suggested long ago~\citep[e.g.][]{rosswog_03}. Recently, \cite{most23_jet} conducted numerical simulations showing that an HMNS can launch mildly-relativistic outflows that may explain precursor emission of Type I GRBs, or in some cases the Type I GRB itself.

\subsubsection{Supramassive Neutron Stars}
\label{sec:SMNS}
\begin{table}
    \centering
    \begin{tabular}{|l|l|l|}
    \hline
    Initial period & $P_0$ & $10^{-3} \, \text{s}$ \\
    Moment of Inertia & $I$ & $10^{45} \, \rm g \, cm^2$ \\
    Polar magnetic field strength & $B_p$ & $10^{15} \, \text{G}$ \\
    Radius & $R$ & $10^{6} \, \text{cm}$ \\
    Efficiency & $\eta_X$ & $0.1$ \\
    Remnant mass & $M_\text{rem}$ & $2.7 \, M_\odot$ \\
    TOV mass limit & $M_{\text{TOV}}$ & $2.15 \,M_\odot$ \\
    \hline
    \end{tabular}
    \caption{Parameters for the model SMNS used in this work.}
    \label{tab:SMNS}
\end{table}

If $M_{\rm rem} \sim (1.0 - 1.2) \, M_{\rm TOV}$, then magnetic braking alone will not induce collapse. A rapidly and rigidly rotating supramassive neutron star (SMNS) can withstand gravitational collapse by rigid rotational energy alone. Once magnetic dipole and gravitational radiation expels sufficient rotational energy, the SMNS will collapse into a BH. The corresponding spindown luminosity is
\begin{equation}
    \label{eq:spindown}
    L_{\rm sd}(t)=\dot E_{\rm rot}(t) = I\Omega\dot\Omega = -\frac{B_p^2 R^6 \Omega^4(t)}{6c^3} -\frac{32 G I \epsilon^2 \Omega^6(t)}{5c^5} 
\end{equation}
where $B_p$ is the polar magnetic field strength, $R$ is the radius, $\Omega$ is the angular velocity, $I$ is the moment of inertia, and $\epsilon$ is the ellipticity. The first term represents magnetic dipole spindown, and the second term represents GW spindown. Note that the so-called ``dipole spindown'' term includes the traditional vacuum dipole spindown component and the magnetospheric wind component, which are of the same order but have different inclination-angle dependencies so that the final expression is insensitive to the inclination angle, see e.g. \citep{goldreich_69, spitz_06}.

For typical SMNS parameters (Table~\ref{tab:SMNS}), magnetic dipole radiation is far more efficient at dissipating rotational energy than gravitational wave radiation. Thus, the spindown luminosity follows
\begin{eqnarray}
L_{\rm sd}(t) \sim L_{\rm sd}(0)\bigg(1+\frac{t}{t_{\rm sd}}\bigg)^{-2},
\label{eq:l_sd}
\end{eqnarray}
where\footnote{As in \cite{gao+16_binary_ns}, one can include both magnetic dipole and GW radiations and derive the spindown law directly from equation~\ref{eq:spindown},
\[
    \label{eq:spindown_law}
    t = \frac{a}{2b^2} \ln\left[\left(\frac{a\Omega_0^2 + b}{a\Omega^2 + b}\right)\frac{\Omega^2}{\Omega_0^2}\right] + \frac{\Omega_0^2 - \Omega^2}{2 b \Omega_0^2 \Omega^2}
\]
where $a = (32 G I\epsilon^2)/(5c^5)$ and $b = (B_p^2 R^6)/(6c^3I)$. }
\begin{eqnarray}
    t_{\rm sd} = \frac{E_{\rm rot}}{\dot{E}_{\rm rot}}=\frac{3 c^3 I}{B_p^2 R^6 \Omega_0^2}.\label{eq:t_sd}
\end{eqnarray}
The collapse time of an SMNS can be approximated by its spindown timescale, corresponding to the time at which the star has lost sufficient rotational support. A detailed calculation of the collapse time depends on the NS EoS, $\Omega_0$ and $B_p$~\citep[see e.g.,][]{beniamini_lu_21}. In this paper, we follow the results of \cite{ai_20_eos} and adopt $M_{\rm TOV} = 2.15 M_\odot$, $M_{\rm SMNS} = 1.25 \, M_{\rm TOV} \approx 2.7 M_\odot$ as a typical model. For an model SMNS with parameters listed in Table~\ref{tab:model_parameters}, the expected collapse time is $t_c\sim t_{
\rm sd} \approx 300 \, \rm s$. 
We note that the model parameters used here are consistent with methods to obtain independent constraints of GW170817/GRB 170817A from the ejecta mass and jet-launching delay timescale~\citep{margalit_17, gill_19}.

\subsubsection{Stable Neutron Star}
\label{sec:stable_ns}

If $M_{\rm rem} < M_{\rm TOV}$, the remnant will evolve as a stable NS that can survive indefinitely. Shortly after merger, there may be a brief phase of differential rotation before settling into uniform rotation. 
Depending on its spin period and magnetic field strength, the remnant may power additional EM signatures---either as sustained emission, akin to radio pulsars, or as episodic bursts, resembling those from soft gamma repeaters and anomalous X-ray pulsars.

\subsection{Viewing geometry}
\label{sec:geometry}

\begin{figure}
    \centering
    \includegraphics[width=0.95\linewidth]{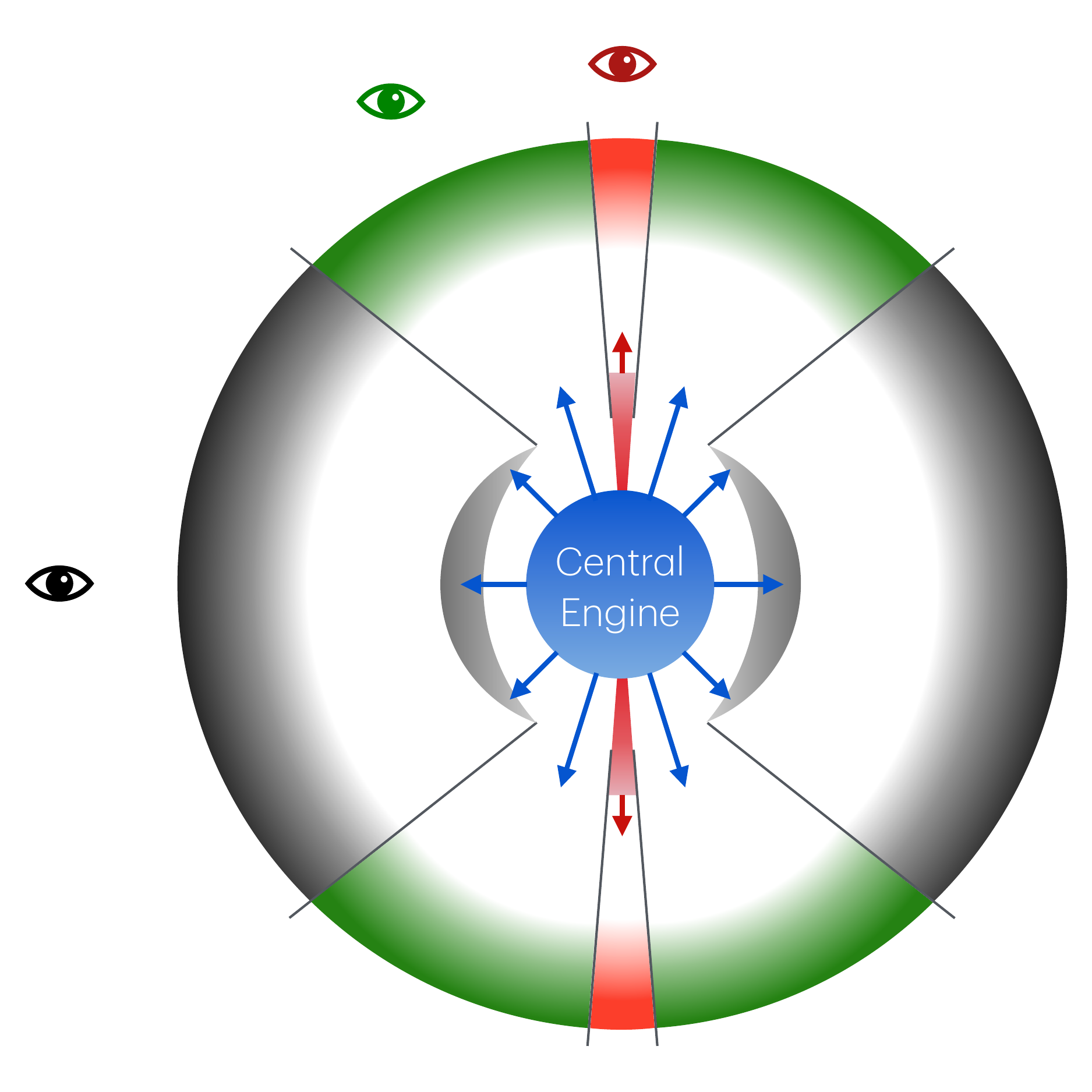}
    \caption{The geometric configuration of a post-merger central engine, which includes a jet zone (red), a free zone (green), and a trapped zone (black). In the jet zone, one sees the brightest region of emission from the jet core. In the free zone, one sees off-axis jet emission and can observe the central engine directly. In the trapped zone, relevant only when considering an opaque ejecta (i.e., a kilonova), one cannot observe the central engine directly but may still see off-axis emissions from the jet and central engine. Adapted from~\cite{sun_17}. 
    }
    \label{fig:viewing_zones}
\end{figure}

In BNS mergers and BH-NS mergers with successful tidal disruption of the neutron star, a fraction of disrupted material could be ejected. This tidal dynamical ejecta is preferentially launched along the equatorial plane, reflecting the orbital angular momentum of the system \citep{kyutoku_15, kawaguchi_15_bhns, lamb_17, kawaguchi_16_bhns, foucart_20, sadeh_24}.

The final jet structure depends on the unknown delay time between the merger and jet launching. 
If one assumes a short jet-launching delay timescale of $\lesssim 0.1 \rm{\, s}$, the jet would quickly penetrate through the ejecta without forming a significant cocoon and evolve into a Gaussian-like jet profile shortly thereafter~\citep{mb_14, geng_19}.
For longer jet-launching delay timescales, the jet-ejecta interaction would generate a slow, mildly-energetic cocoon that surrounds the central jet, making a two-component jet structure~\citep{gottlieb_18, salafia_20}.
In general, regardless of the central engine, the observed X-ray emission is strongly influenced by geometry of the system (see Fig.~\ref{fig:viewing_zones}).

If the observer's LoS is closely aligned with the jet axis---i.e., the viewing angle $\theta_v$ is smaller than the jet opening angle $\theta_c$ (the \textit{Jet zone})---bright gamma-ray and X-ray prompt emission from the jet is beamed directly toward the observer. For viewing angles slightly outside the jet ($\theta_v > \theta_c$) but not significantly obstructed by any dynamical ejecta (the \textit{Free zone}), off-axis prompt gamma-ray/X-ray emission may still be detected, depending on the angular structure of the jet and emission from the central engine. For magnetar engines with extended X-ray emission powered by a spindown wind, the emission may be directly observed from both the Jet and Free zones. If the observer's LoS lies within the \textit{Trapped zone}, where the central engine is obscured by an optically thick ejecta, any X-ray signals from the central engine would be significantly suppressed.

The optical depth of the dynamical ejecta is given by
\begin{equation}
\begin{aligned}
  \tau &= \kappa \rho \Delta r \sim \kappa \frac{3M_\text{ej}}{4 \pi (vt)^2} \\
  & \approx 6 \times 10^7 \left(\frac{\kappa}{1 \text{ cm}^2\text{ g}^{-1}}\right)\left(\frac{M}{10^{-2}M_\odot}\right) \\ &\left(\frac{v}{0.1c}\right)^{-2} \left(\frac{t}{100\text{ s}}\right)^{-2}  
\end{aligned}
\end{equation}
where $\rho$ is the average ejecta density and $\Delta r\sim vt$ is the ejecta shell width. 
Initially, the ejecta is optically thick, however, as the ejecta expands, the optical depth decreases, and the \textit{Trapped zone} eventually becomes transparent to X-rays. The corresponding transparency time is
\begin{equation}
\begin{aligned}
    t_\text{thin} &= \sqrt{\frac{3\kappa M_\text{ej}}{4\pi \beta v^2}} \\ &\approx 1.33 \times 10^5 \left(\frac{\kappa}{1 \text{ cm}^2\text{ g}^{-1}}\right)^{1/2}  \\ &\left(\frac{M}{10^{-2}M_\odot}\right)^{1/2} \left(\frac{v}{0.1c}\right)^{-1/2}  \text{ s} 
    \end{aligned}
\label{eq:t_thin}
\end{equation}
where $\beta \approx 3$ is a parameter delineating the ejecta density profile \citep{metzger+10_kilonova}. 
Eq.~\ref{eq:t_thin} is based on the assumption that the density profile is uniform.

Central engine activity during $t>t_{\rm thin}$ can be directly observed along any LoS. In contrast, any emission launched earlier than $t_{\rm thin}$ must diffuse through the ejecta, and will be delayed by the diffusion time,
\begin{equation}
    t_\text{diff} = \frac{\Delta r}{c}\tau = \kappa \frac{3M_\text{ej}}{4\pi c v t}.
\end{equation}
Still, during $t < t_{\rm thin}$, central engine activity can influence the dynamics and evolution of the ejecta and associated X-ray signals.
For a magnetar central engine, spindown would inject energy into the ejecta; if the injected energy exceeds the kinetic energy of the dynamical ejecta, the ejecta will be accelerated and significantly reduce $t_{\rm thin}$ (see, e.g., the “mergernova” scenario; \citealt{yu_13,ai_22,ai_25}).
Additionally, the magnetar wind would interact with the dynamical ejecta and drive a shock that can radiate in soft X-rays~\citep{siegel_ciolfi_16_1, siegel_ciolfi_16_2}. In our discussion, the \textit{Trapped zone} refers to the optically thick regions where a ``naked" magnetar wind may not be directly observed.

\section{X-ray Emissions from NS mergers}
\label{sec:emissions}

\subsection{Prompt jet}
\label{sec:prompt}

For both BNS and NS-BH mergers, a relativistic jet may be launched perpendicular to the orbital plane. Energy dissipation within the relativistic jet powers GRB prompt emission. The radiation mechanism of prompt emission is still not fully understood; leading candidates include synchrotron radiation from an optically thin region and quasi-thermal, Comptonized emission near the photosphere. Synchrotron self-Compton, external inverse Compton, and hadronic cascades have been also proposed to contribute to prompt emission (for comprehensive reviews of proposed mechanisms, see, e.g., \citet{meszaros_06,kumar_zhang14_grb,zhang18_grb_book}).
Here, instead of involving uncertainties to model GRB prompt emission beginning from fundamental principles, we opt for a phenomenological approach. The framework described in this section and used to produce prompt X-ray light curves and spectra is packaged as the open-source Python package \texttt{PromptX}~\citep{promptx}. 

We model the temporal profile of a GRB pulse with the fast-rise-exponential-decay (FRED) function~\citep{norris_05, hakkila_08},
\begin{equation}
\label{eq:fred}
    I(t) = \frac{A\lambda}{\exp \left(\tau_1/t + t / \tau_2 \right)},\quad t > 0,
\end{equation}
where $A$ is the normalization constant, $\tau_1$ and $\tau_2$ are the rise and decay timescales, respectively, and $\lambda=\exp(2\mu)$, where $\mu = \sqrt{\tau_1/\tau_2}$. Within this framework, the peak time is $t_{\rm peak} = \sqrt{\tau_1 \tau_2}$, the interval measured between two $1/e^3$ amplitudes is $\Delta t = \sqrt{9 + 12\mu}$, and the pulse asymmetry is defined by $\Delta t/(3 + 2\mu)$. 

We model the GRB spectrum with a Band function~\citep{band_93}
\begin{equation}
\label{eq:band}
N(E) = A \times
\left\{
    \begin{aligned}
   &E^{\alpha} \exp\left(\frac{E \, (\beta - \alpha)}{E_\text{b}}\right), & \text{for } E < E_{\text{b}}, \\
   &E_{\text{b}}^{\alpha - \beta} \exp(\beta - \alpha) \, E^\beta, & \text{for } E \geq E_{\text{b}},
    \end{aligned}
\right.
\end{equation}
where $A$ is the normalization constant, $E_b = (\alpha - \beta) \, [E_p / (2 + \alpha)]$ is the break energy, where $E_p$ is the peak energy in the energy spectrum, and $\alpha$ and $\beta$ are the low- and high-energy photon indices. 

The intrinsic gamma-ray energy profile of the jet is defined as
\begin{equation}
    \epsilon_\gamma (\theta) \equiv \frac{dE_\gamma(\theta)}{d\Omega}.
    \label{eq:eps}
\end{equation}
A cutoff angle, $\theta_{\rm cut}$, is applied when modeling a source with an opaque ejecta (see Fig.~\ref{fig:viewing_zones}).
Typically, one uses one of the following functional forms:
1) A tophat
\begin{equation}
\label{}
\epsilon_\gamma(\theta) = 
\begin{cases}
\epsilon_{\gamma,\,0}, & \theta \leq \theta_{\mathrm{cut}}, \\
0, & \theta > \theta_{\mathrm{cut}},
\end{cases}
\end{equation}
2) A power-law
\begin{equation}
\epsilon_\gamma(\theta) =
\begin{cases}
\epsilon_{\gamma,0}, & \theta \leq \theta_c, \\[0.75em]
\epsilon_{\gamma,0}\,\left( \dfrac{\theta}{\theta_c} \right)^{-k}, & \theta_c < \theta \leq \theta_{\mathrm{cut}}, \\[0.75em]
0, & \theta > \theta_{\mathrm{cut}},
\end{cases}
\end{equation}
where $k$ is the power-law decay index,
3) A Gaussian
\begin{equation}
\label{eq:eps_profile}
\epsilon_\gamma(\theta) = 
\begin{cases}
\epsilon_{\gamma,\,0}  \exp\left(-\theta^2 / 2\theta_c^2\right), & \theta \leq \theta_{\mathrm{cut}}, \\
0, & \theta > \theta_{\mathrm{cut}},
\end{cases}
\end{equation}
where $\epsilon_{\gamma, \, 0}$ is the peak energy per solid angle in gamma-rays and $\theta_c$ is the characteristic Gaussian width of the energy profile.
In this work we adopt a Gaussian for the jet energy profile. We adopt a fixed cutoff angle, $\theta_{\rm cut}$, independent of radius.
We also define an identical counter-jet opposite of the main GRB jet.

The Lorentz factor of the prompt-emitting material is defined by the particular emission mechanism(s) responsible for prompt emission. Here, we adopt an observational correlation between the the observed $\gamma$-ray isotropic energy and the initial Lorentz factor derived from GRB afterglow observations\footnote{To derive $\Gamma$, one makes use of the observed afterglow peak time and the blastwave deceleration model of \cite{rees_meszaros_92} and \cite{sari_piran_99}.}~\citep{liang+10_relation, ghirlanda+11_relation},
\begin{equation}
\label{eq:lg_relation}
    \Gamma(E_{\gamma, \, \rm iso}) = \Gamma_0 \,(E_{\gamma, \, \rm iso} / 10^{52} \, \rm{erg})^{1/4} + 1,
\end{equation}
where $\Gamma_0\sim 180$. \cite{ghirlanda_18} revisited the relation using an updated sample of GRBs and found similar scalings.
The samples in which this empirical correlation is satisfied spread a few orders of magnitude in $E_{\rm iso}$, which can be attributed to a wide range of viewing angles under the quasi-universal jet assumption. Under this assumption, all GRB jets share a similar angular energy and Lorentz factor distribution, with observed differences primarily arising from the observer's viewing angle rather than intrinsic variations in the jet itself~\citep{zhang_02, lloyd_ronning_04_quasi}. Thus, the $\Gamma-E_{\rm \gamma,iso}$ relation can be extended to off-axis events and provides a framework grounded in observables to interpret both on- and off-axis GRB observations. 

As discussed further in Section~\ref{sec:results}, we adopt a jet radiative efficiency for prompt emission, $\eta_\gamma = 0.01$, consistent with observational constraints~\citep{beniamini_15, beniamini_16_efficiciency, salafia_21_efficiency} to estimate the Lorentz factor of prompt-emitting material as $\Gamma_{0, \gamma} = \Gamma_0 / (1-\eta_\gamma) \approx 180$~\citep{zhang_21}.

Consider a structured GRB jet with two viewing positions: one along the jet axis and one at an angle $\theta_v$. The on-axis observer sees the luminous jet core, whereas the off-axis observer sees only the dimmer outer jet material. In both cases, emission from surrounding angular regions contributes to the observed signal, weighted by relativistic Doppler beaming.

The Doppler factor can generally be written as
\begin{equation}
    \label{doppf}
    \mathcal{D} \equiv \frac{1}{\Gamma(1-\beta\cos\theta_v)}.
\end{equation}

The on- and off-axis observers of a structured GRB jet will each calculate a particular Doppler factor, $\mathcal{D}_{\rm on}$ and $\mathcal{D}_{\rm off}$, respectively. The ratio of Doppler factors is then $\mathcal{R_\mathcal{D}} \equiv \mathcal{D_{\rm off}}/ \mathcal{D_{\rm on}} < 1$. More generally, an observer with an arbitrary LoS $(\theta_v, \phi_v)$, will calculate a unique $\mathcal{R}_\mathcal{D}$ for every other $(\theta_i, \phi_j)$ patch, 
\begin{equation}
    \label{eq:rdopp}
\mathcal{R_\mathcal{D}} (\theta_v,\phi_v,\theta_i,\phi_j) = \frac{\mathcal{D}(\theta_v, \phi_v)}{ \mathcal{D}(\theta_i, \phi_j)} = \frac{1-\beta}{1-\beta\cos\theta_w} < 1,
\end{equation}
where 
\begin{equation}
\label{eq:angular_d}
    \theta_w=\cos^{-1} \left[ \sin \theta_v \sin \theta_i \cos(\phi_v - \phi_j) + \cos \theta_v \cos \theta_i \right]
\end{equation}
is the angular distance between $(\theta_v,\phi_v)$ and $(\theta_i, \phi_j)$. 
$\mathcal{R}_\mathcal{D}$ relates observed jet properties between the on- and off-axis observers,
\begin{align}
    dt_\text{off} &=\mathcal{R_\mathcal{D}}^{-1} \, dt_\text{on}, \\
    \epsilon_\text{off} &= \mathcal{R_\mathcal{D}}^3 \,\epsilon_\text{on}.
\end{align}

\begin{figure}
    \centering
    \includegraphics[width=\linewidth]{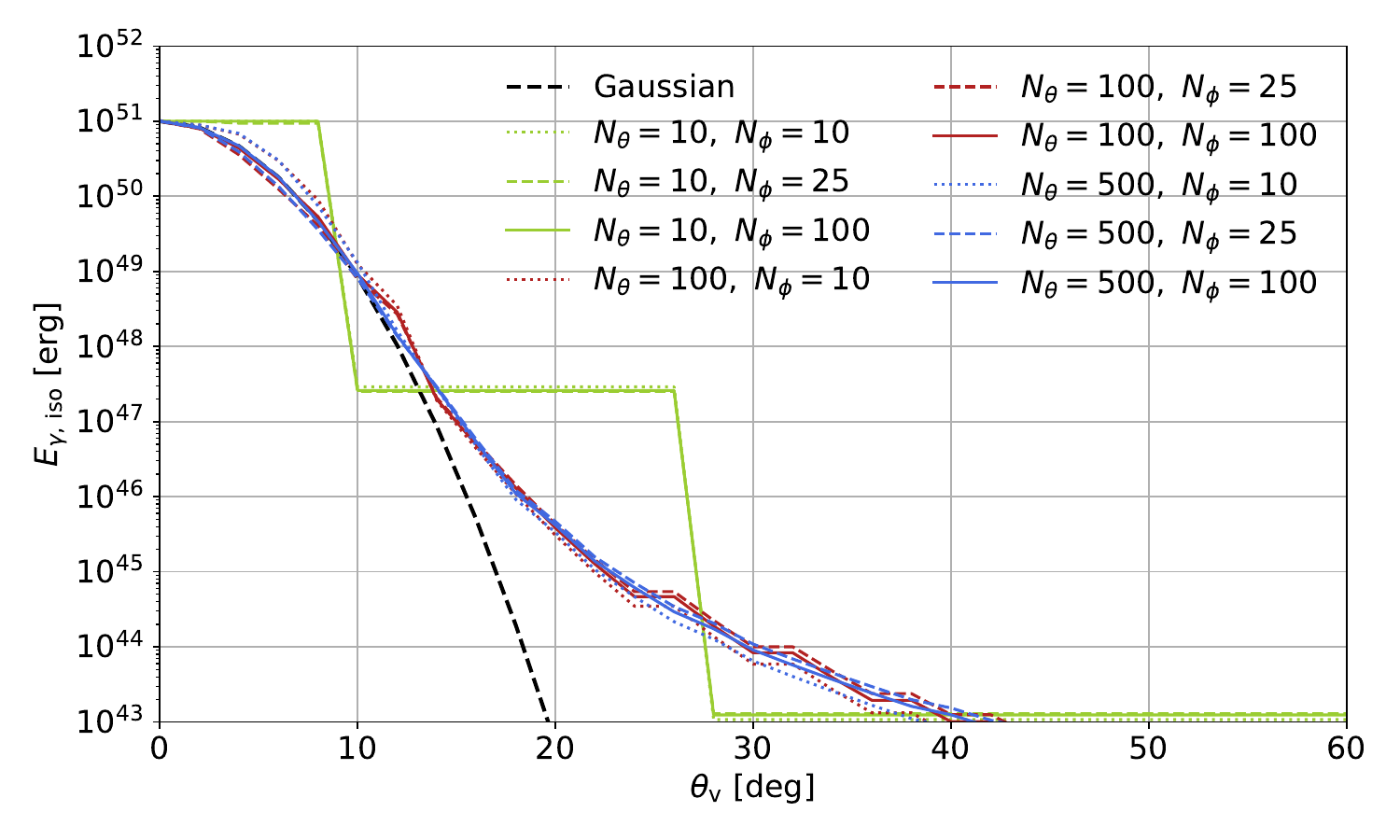}
    \caption{Observed isotropic-equivalent energy, $E_{\gamma, \, \rm iso} = 4\pi\bar\epsilon_\gamma$ (Eq.~\ref{eq:eps_bar}), for an observer at various $\theta_v$.} 
\label{fig:convergence}
\end{figure}

Numerically, we employ a $500 \times 100$ $(\theta, \phi)$ grid uniformly spaced in $\phi$ and $\cos\theta$, providing finer sampling near the jet axis. This choice of resolution ensures an optimal balance between performance and accuracy (Fig.~\ref{fig:convergence}).
We define a Gaussian GRB jet with intrinsic $\epsilon_\gamma$ (Eq.~\ref{eq:eps_profile}) and $\Gamma$ (Eq.~\ref{eq:lg_relation}) profiles. In the Trapped zone, $\epsilon_{\gamma}(\theta > \theta_{\rm cut}) = 0$ and $\Gamma(\theta > \theta_{\rm cut}) = 1$. 

\begin{figure*}
    \centering
    \includegraphics[width=0.328\linewidth]{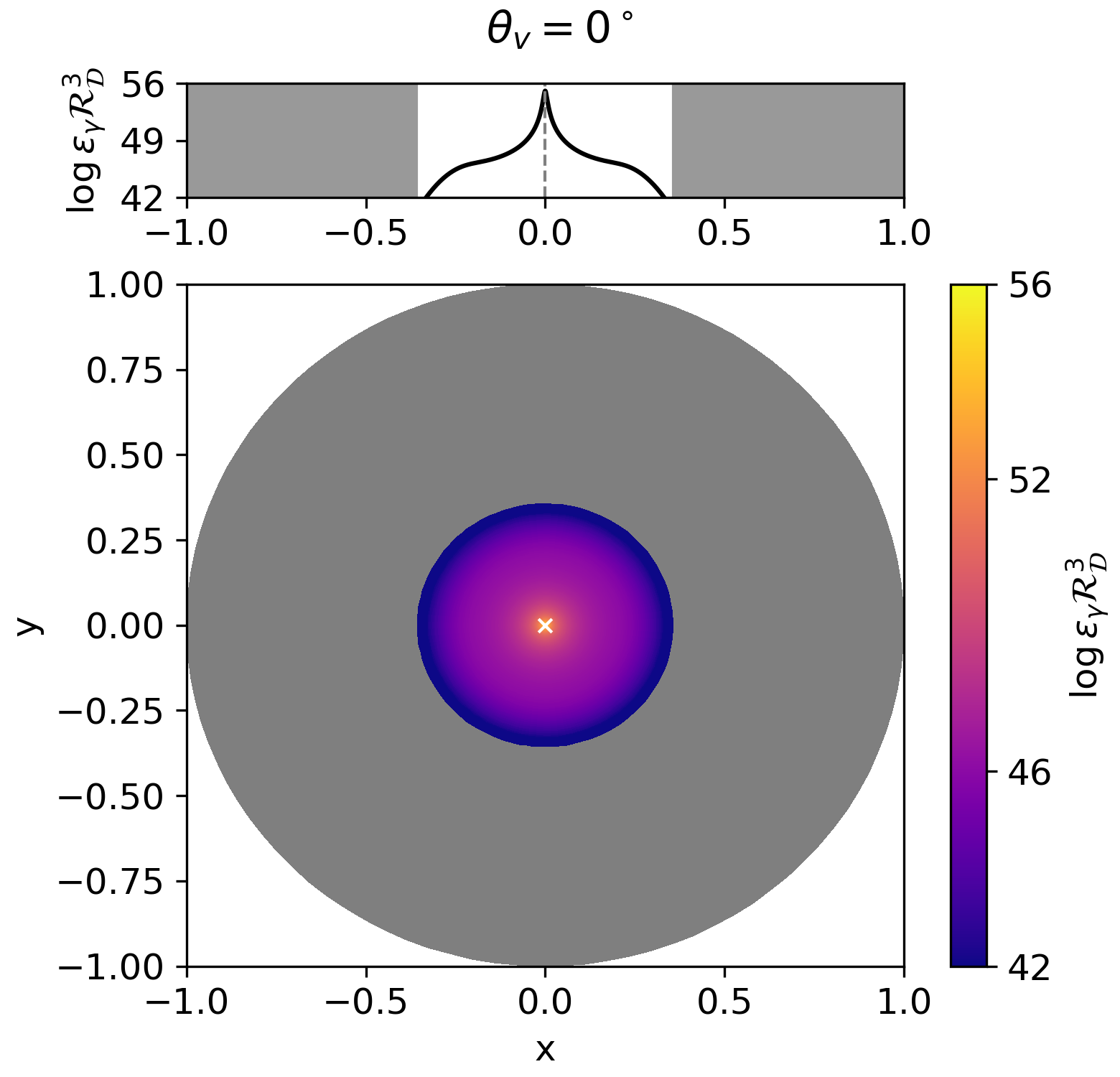}
    \includegraphics[width=0.328\linewidth]{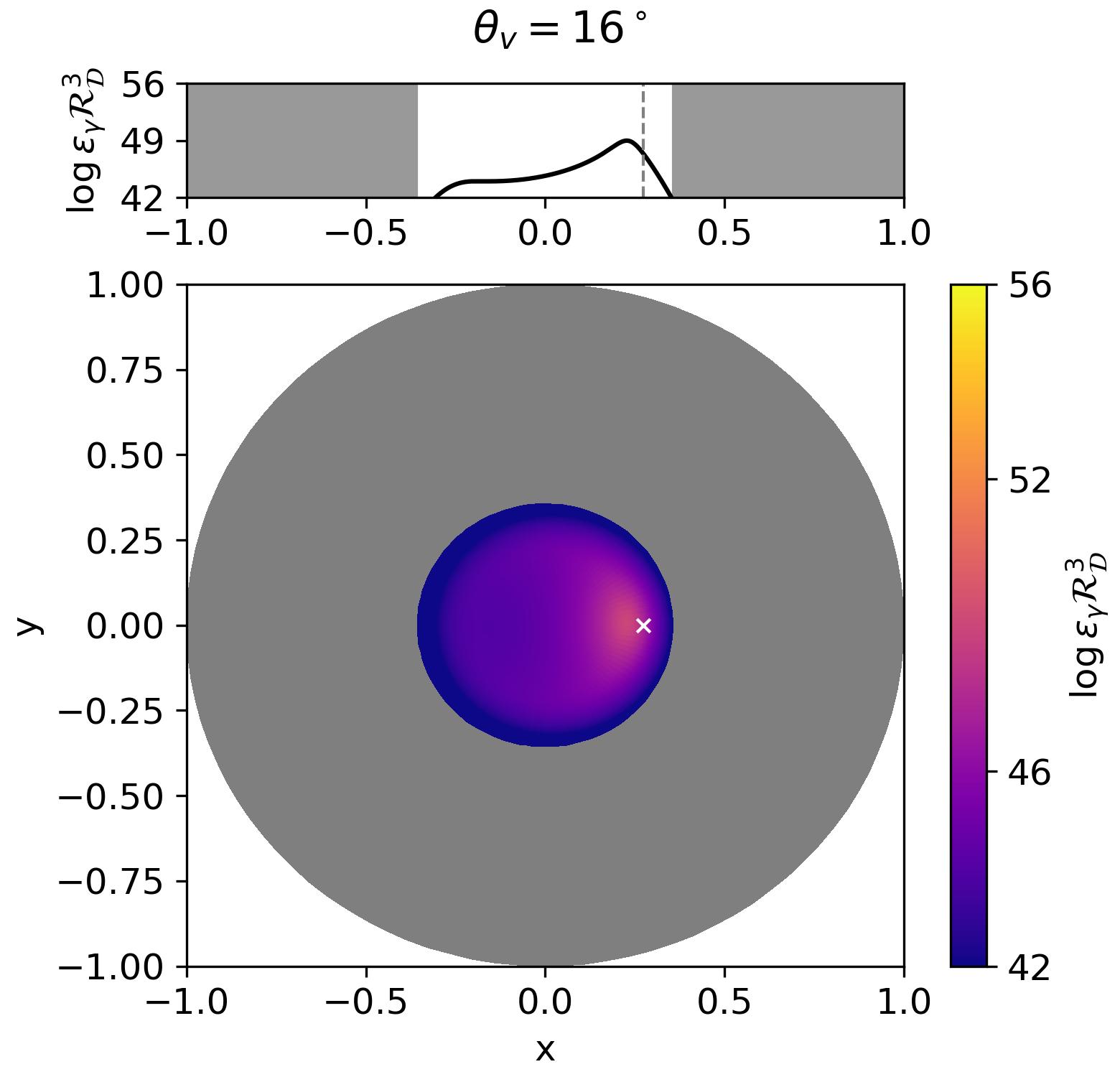}
    \includegraphics[width=0.328\linewidth]{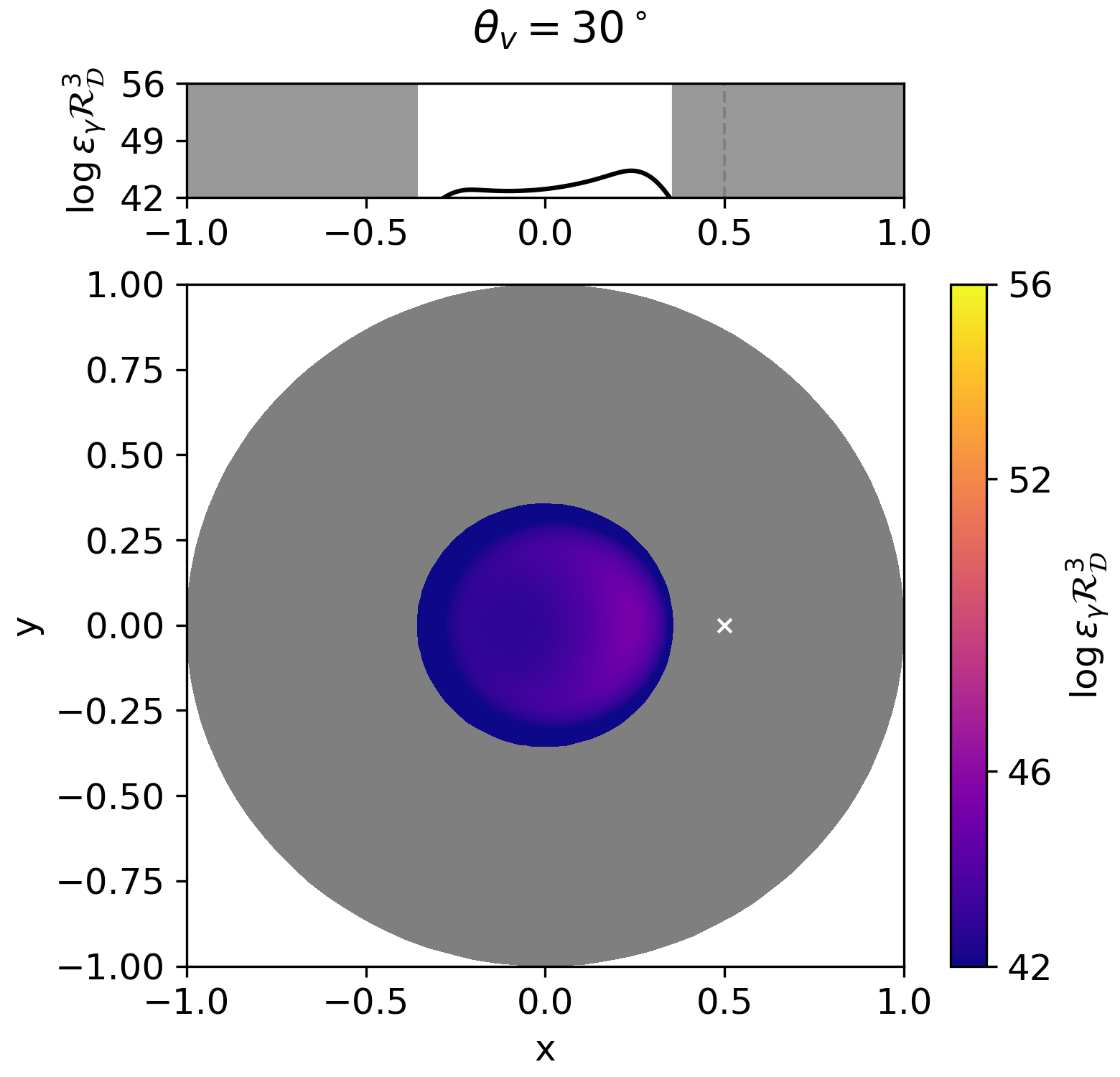}
    \includegraphics[width=0.328\linewidth]{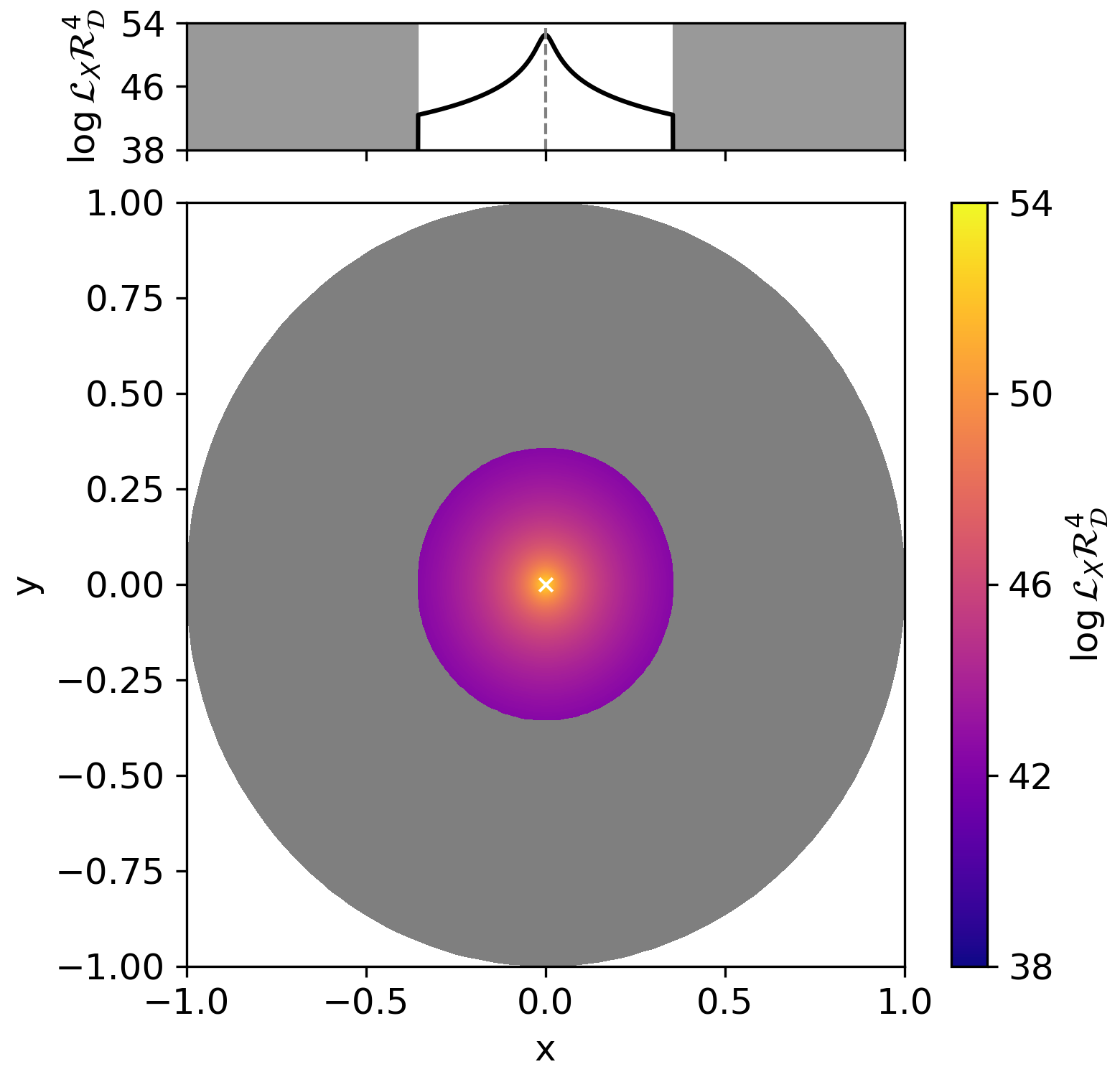}
    \includegraphics[width=0.328\linewidth]{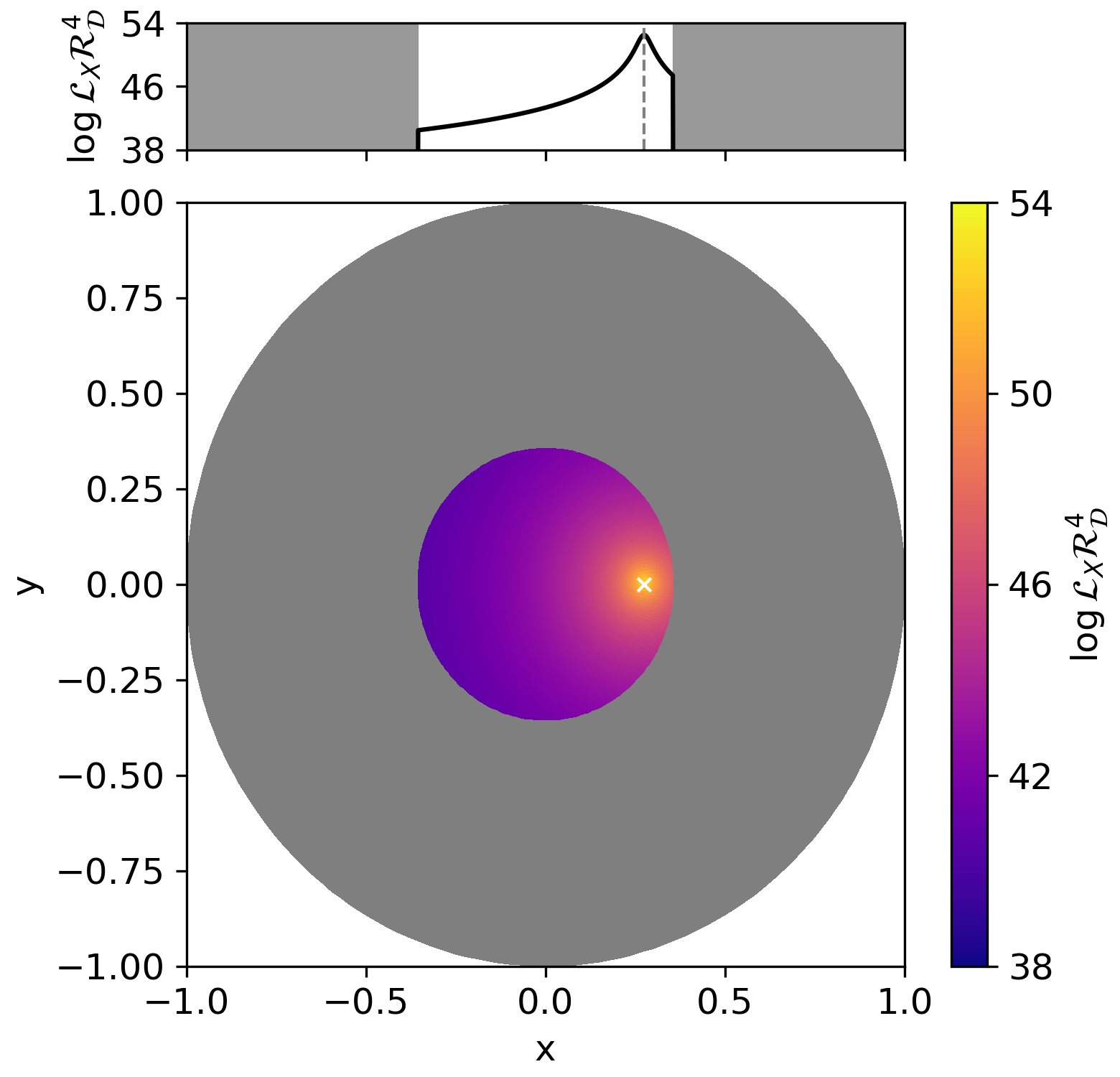}
    \includegraphics[width=0.328\linewidth]{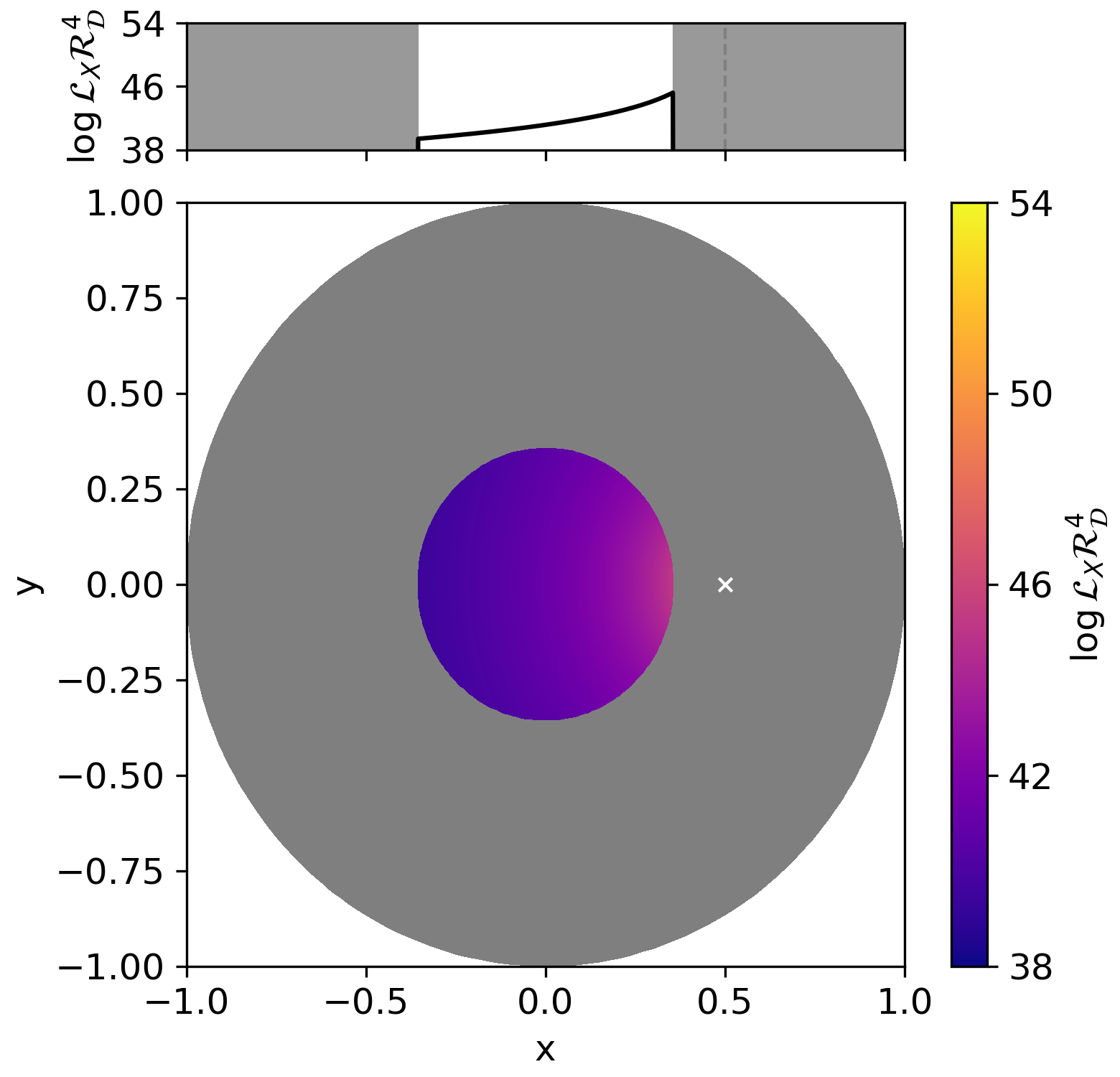}
    \caption{Maps of observed energy per solid angle, $\epsilon_\gamma\mathcal{R}_\mathcal{D}^3$, for a $\theta_c = 3^\circ$ Gaussian GRB jet (top) and isotropic wind (bottom) for an observer at $\theta_v=0^\circ$ (left), $\theta_v=16^\circ$ (middle), and $\theta_v=40^\circ$ (right), with the observer's position marked by the white cross. The gray region indicates the Trapped zone, where $\theta > \theta_{\rm cut} = 20^\circ$. The upper panels show 1D meridional slices through $\theta_v$ (vertical dashed line).}
\label{fig:doppler}
\end{figure*}

Consider again the off-axis observer at $(\theta_v, \phi_v)$. The energy per solid angle of the $(\theta_i, \phi_j)$ patch would appear Doppler-shifted, $ \epsilon_{\gamma}(\theta_i, \phi_j) \mathcal{R}_\mathcal{D}^3$. Fig.~\ref{fig:doppler} maps the observed energy per solid angle for various LoS.
The \textit{effective} energy per solid angle that an observer would detect from emitting regions ($\theta < \theta_{\rm cut}$) is
\begin{equation}
\label{eq:eps_bar}
    \bar\epsilon_\gamma (\theta_v, \phi_v) = k_0 \frac{\sum_{i,j} \epsilon_{\gamma}(\theta_i, \phi_j) \mathcal{R}_\mathcal{D}^3 \Delta\Omega_{i,j}}{\sum_{i,j} \Delta\Omega_{i,j}},
\end{equation}
where $k_0$ is the normalization factor, defined such that an on-axis observer detects an isotropic-equivalent energy in gamma-rays, $4\pi \bar\epsilon_{\gamma}(0, 0)$, equal a given observed $E_{\rm \gamma, \, iso}$.

At each $(\theta_i,\phi_j)$ patch, one can define an on-patch observer-frame spectrum per solid angle, ${\partial N_{i,j}(E|E_p)}/{\partial\Omega}=\mathcal{N}_{i,j}(E|E_p)$ (e.g., a Band function, Eq.~\ref{eq:band}). The notation $(E|E_p)$ indicates that observed energy, $E$, is not necessarily Doppler-shifted but indirectly scaled through $E_p$, which is defined according to the empirical Amati relation, $E_p \propto E_{\gamma, \, \rm iso}^{a}$, where $E_{\gamma, \, \rm iso} = 4\pi\bar\epsilon_\gamma(\theta_w)$.
Classically, for Type II GRBs, $a = 0.5$~\citep{amati_02}. Later, it was found that Type I GRBs also have a similar correlation but with a track above the Type II branch~\citep{zhang_09}. 
\cite{minaev_20} analyzed an updated sample of events and found $a = 0.4$ for short (Type I GRBs), which is adopted in this work. 

\begin{figure}
    \centering
    \includegraphics[width=\linewidth]{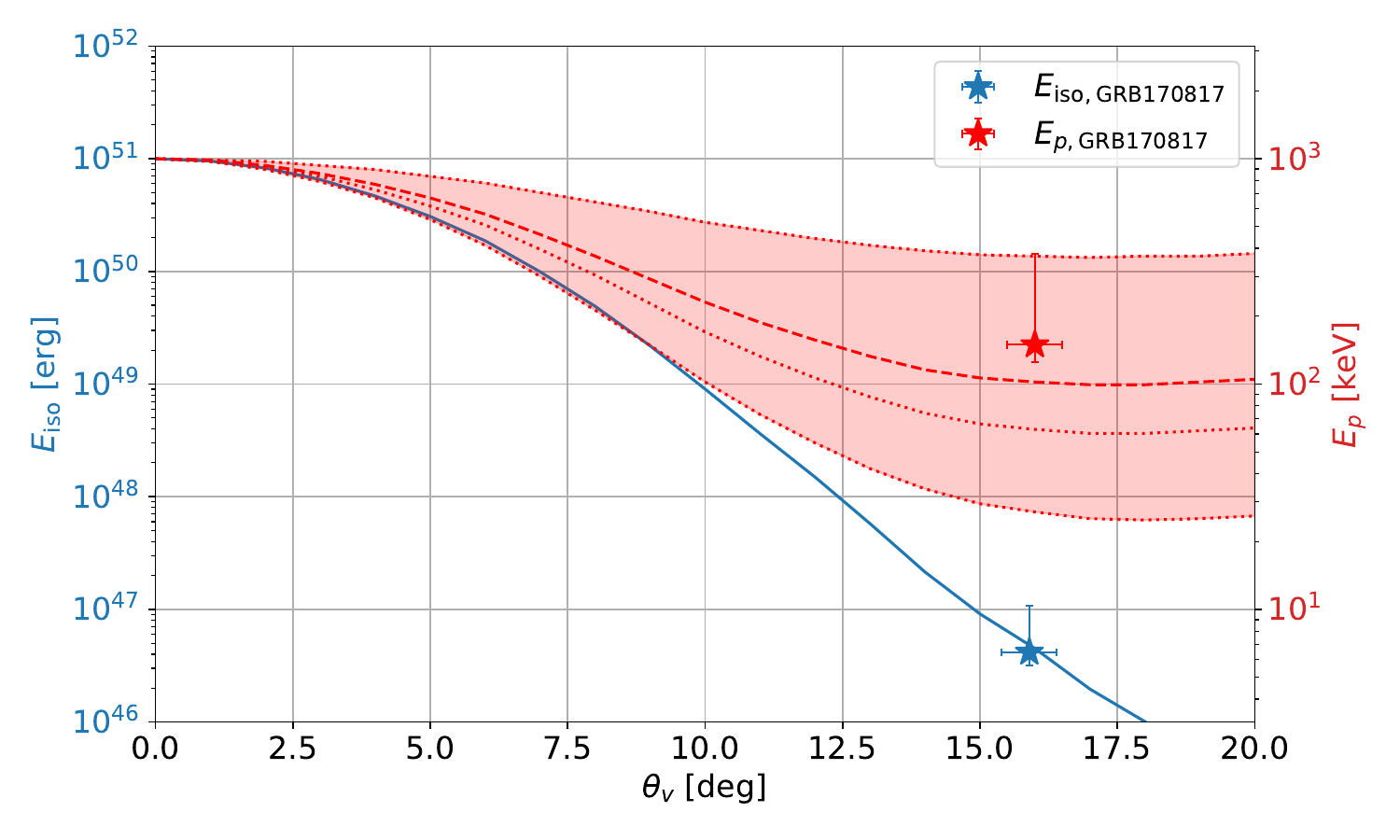}
    \caption{Observed isotropic-equivalent energy, $E_{\gamma, \rm iso}$ (blue, left axis) and observed spectral peak energy $E_p$ (red, right axis) as a function of viewing angle $\theta_v$. The red dashed line bounded by the red shaded region shows the Amati relation derived by \cite{zhang_09}, $E_p\propto E_{\rm iso}^{0.34 \pm 0.19}$. The red solid line shows the Amati relation derived by \cite{minaev_20} for Type I GRBs, $E_p\propto E_{\rm iso}^{0.4}$. Deviations at larger $\theta_v$ occur because the observed spectrum is dominated by emission from the jet core, which possesses a higher intrinsic $E_p \propto \mathcal{R}_\mathcal{D}$ despite strong Doppler deboosting of $E_{\rm iso} \propto \mathcal{R}_\mathcal{D}^3$. GRB 170817A data obtained from~\citep{zhang_18}.
    }
    \label{fig:amati}
\end{figure}

As we will demonstrate later, assigning a Band spectrum to each grid point yields observational properties consistent with the Amati relation near $\theta_v \sim \theta_c$---the regime wherein the relation was originally established---but deviates at larger $\theta_v$. In the case of GRB 170817A, which exhibited unusually low luminosity and soft spectral characteristics consistent with an off-axis viewing angle~\citep{mooley_18, troja_19, 170817}, this effect remains broadly consistent with observations within uncertainties (Fig.~\ref{fig:amati}).

We adopt typical values for Type I GRBs: $E_{\rm p, \, 0}\sim 1 \, \rm MeV$, $\alpha=-1$, and $\beta = -2.3$. For each spectrum, the total gamma-ray energy per solid angle is defined as
\begin{equation}
\label{eq:e_tot}
    \epsilon_{\gamma,i,j} = A\int_{E_1}^{E_2} E\mathcal{N}_{i,j}(E|E_{\rm p,0}) \, dE,
\end{equation}
where $E_1$ and $E_2$ are the energy bounds of a particular gamma-ray detector (e.g. $E_1=10 \, \rm keV$ and $E_2=1000 \, \rm keV$ for Fermi/GBM), and $A$ is the normalization factor.
For each normalized Band function, the spectrum can be extended to X-rays and integrated to obtain the total X-ray energy per solid angle in the corresponding patch, $\epsilon_{X,i,j} = A \int_{E_1}^{E_2} E\mathcal{N}_{i,j}(E|E_{\rm p,0}) \, dE$, where $E_1=0.3 \, \rm keV$ and $E_2 = 10 \, \rm keV$.
We note here that while the Band function prescribes a single power-law index for the low-energy tail that could, in principle, be extrapolated to X-rays, several studies have found evidence of additional sub-MeV spectral breaks in Type II GRBs~\citep{oganesyan_17, ravasio_18_2sbpl, ravasio_19_2sbpl, toffano_21}, even though these low-energy spectral breaks have not been robustly identified in Type I GRBs~\citep{ravasio_19_2sbpl, toffano_21}. Since the intrinsic spectrum of GRBs is still unknown, here we adopt the typical Band function as a proof-of-concept model. 

Then, one can model an intrinsic gamma-ray (or X-ray) luminosity light curve per solid angle $\mathcal{L}_{i,j}={\partial L_{i,j}}/{\partial\Omega}$ at each patch (e.g., a FRED profile; Eq.~\ref{eq:fred}) normalized such that the fluence equals the total gamma-ray (or X-ray) energy per solid angle:
\begin{align}
    \epsilon_{\gamma,\, i, j} &= \int_{t_1}^{t_2} {\mathcal{L}_{\gamma,\, i, j}(t)} \, dt, \\ 
    \epsilon_{X,\, i, j} &= \int_{t_1}^{t_2} {\mathcal{L}_{X,\, i, j}(t)} \, dt.
\end{align}
As a case study of GRB170817A, we adopt the FRED pulse parameters, $\tau_1 = 0.02 \, \rm s$ and $\tau_2 = 0.2 \, \rm s$, so that the observed Doppler-broadened $T_{90}$ matches the Fermi/GBM measurement of GRB170817A, $T_{90, \, \rm GRB \, 170817A} = (2 \pm 0.5) \, \rm s$, assuming a typical fluence sensitivity of $~10^{-7} \, \rm erg \, cm^{-2}$~\citep{grb170817_fermi}. 
For $E_{\gamma,\,\rm iso,\,GRB\,170817A} = 4.17^{+6.54}_{-0.99} \times10^{46}\,\rm erg$~\citep{zhang_18}, the isotropic-equivalent luminosity in gamma-rays is $L_{\gamma , \, \rm iso, \,GRB\,170817A} = 2.09^{+3.27}_{-0.50} \times 10^{46} \,\rm erg\,s^{-1}$. 

\begin{figure}
    \centering
    \includegraphics[width=\linewidth]{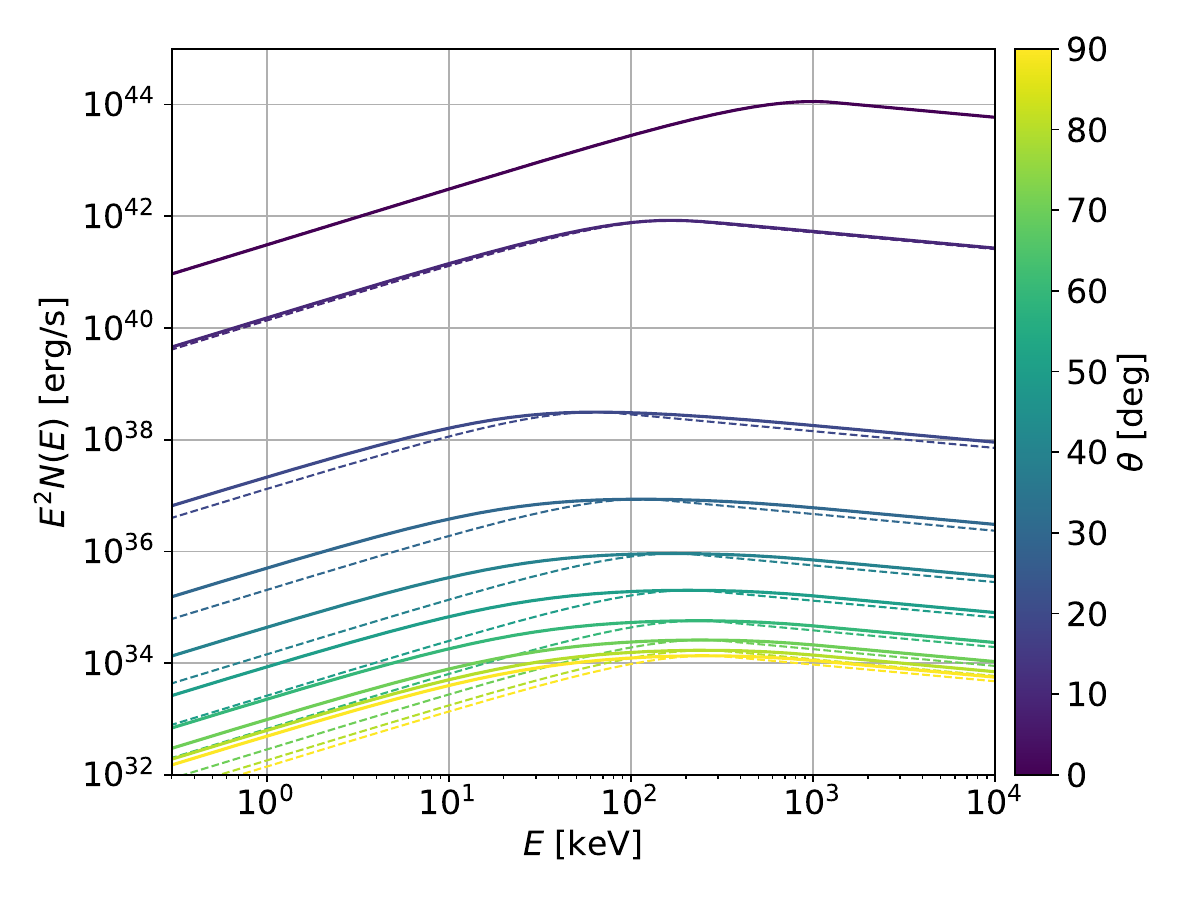}
    \caption{Total observed spectra (solid) compared to reference Band functions (dashed) with $\alpha = -1$ and $\beta = -2.3$ for various $\theta_v$.}
    \label{fig:convergence_spec}
\end{figure}

For the observer at $(\theta_v,\phi_v)$, the observed spectra per solid angle of each $(\theta_i, \phi_j)$ patch is Doppler-shifted,
\begin{equation}
\label{eq:spec_obs}
    \mathcal{N}_{i, j,\rm obs}(E) = \mathcal{N}_{i,j}(E|E_p(\theta_w)) \, \mathcal{R}_\mathcal{D}(\theta_{v}, \phi_{v}, \theta_i, \phi_j).
\end{equation}
One obtains the total observed spectrum per solid angle by combining the contributions from all emitting patches,
\begin{equation}
    \mathcal{N}_{\rm tot}(E) = \frac{\sum_{i,j} \mathcal{N}_{i,j, \,\rm obs}(E)\Delta\Omega_{i,j}} {\sum_{i,j}  \Delta\Omega_{i,j}}.
    \label{eq:N_E_tot}\\
\end{equation}
For intrinsic spectra modeled as a Band function, the total observed spectrum still resembles a Band function, but with a smoother break, as shown in Fig.~\ref{fig:convergence_spec}.

For an observer at $(\theta_{\rm v},\phi_{\rm v})$, we transform the emission from each $(\theta_i,\phi_j)$ surface element from the source (engine) frame to the observer frame. We denote the emission time at the central engine by $t_{\rm eng}$ and the photon arrival time measured by a distant observer by $t_{\rm obs}$.

The emitting radius of each patch is computed as
\begin{equation}
    R_{{\rm em} ,i,j}(t) = \frac{\beta_{i,j} c\, t_{\rm eng}}{1-\beta_{i,j}},
\end{equation}
where $\beta_{i,j} = \sqrt{1 - 1 / \Gamma_{i,j}^2}$. For a finite-duration pulse in a structured outflow, the emission radius varies with the source-frame emission time.

The corresponding observer time of each patch follows as
\begin{equation}\label{eq:em-r}
    t_{{\rm obs}, i,j}
    =
    t_{\rm eng, i,j}
    +
    \frac{R_{{\rm em},i,j}}{c}
    \left(1-\cos\theta_{\rm w}\right),
\end{equation}
where $\theta_{\rm w}$ is the angular separation between the $(\theta_i,\phi_j)$ patch and the observer's line of sight (Eq.~\ref{eq:angular_d}). This mapping inherently accounts for the light-travel time of photons emitted from different radial and angular extents of the outflow.

\begin{figure*}
    \centering
    \includegraphics[width=0.328\linewidth]{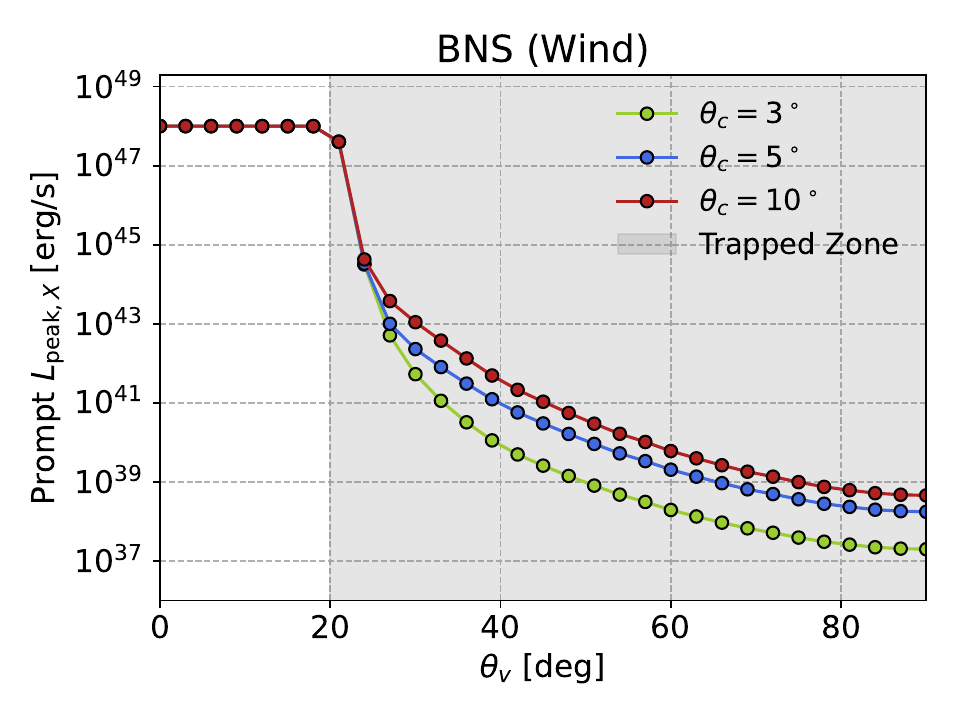}
    \hspace*{-0.5em}
    \includegraphics[width=0.328\linewidth]{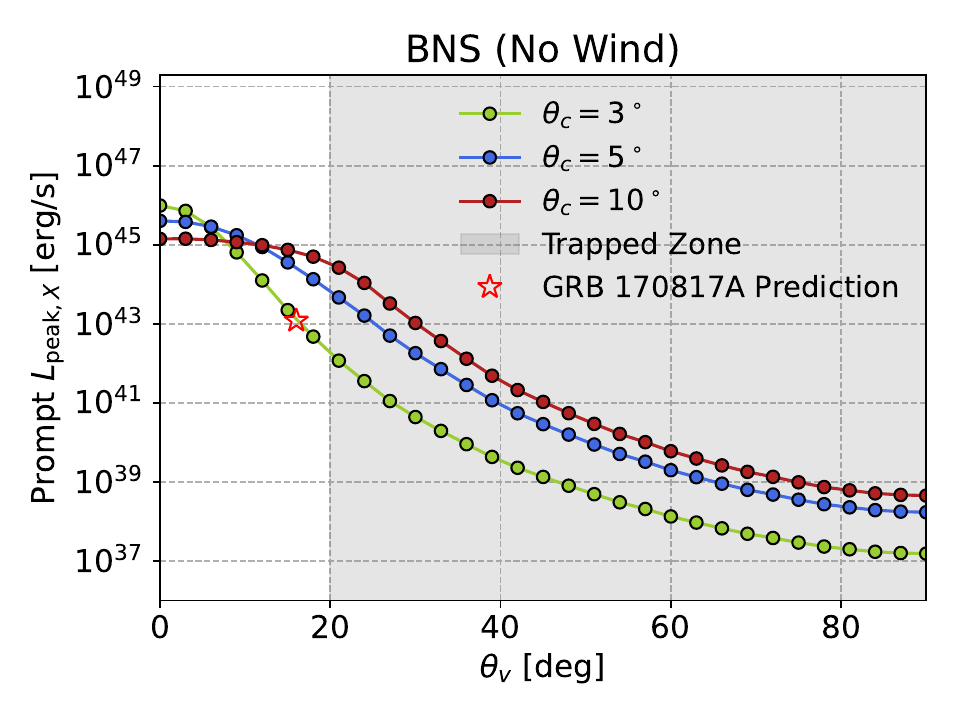}
    \hspace*{-0.5em}
    \includegraphics[width=0.328\linewidth]{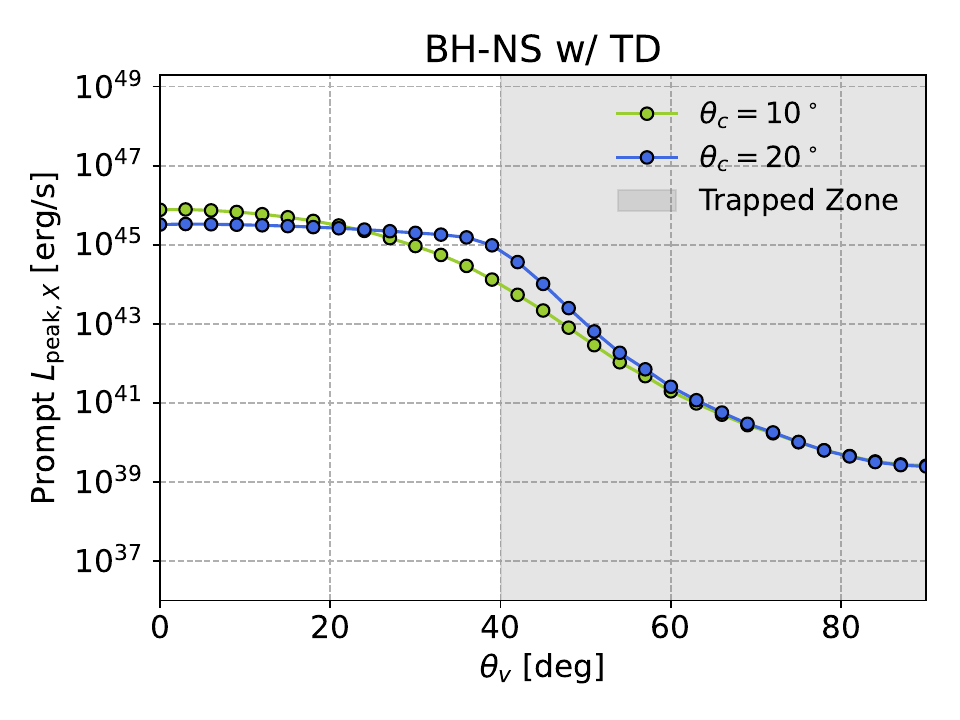}
    \caption{Prompt X-ray emission peak luminosity as a function of viewing angle. \textit{Left panel} is for BNS mergers with a magnetar central engine wind component (i.e., BNS-I and early-time BNS-II). \textit{Middle panel} is for BNS mergers without a wind component (i.e., BNS-III/BNS-IV and late-time BNS-II). \textit{Right panel} is for BH-NS mergers with tidal disruption. Total energy is fixed such that $E_{\gamma, \, \rm iso} = 10^{51} \, \rm erg \,$ for $\theta_c = 3^\circ$. The hollow red star in the middle panel is the predicted peak prompt X-ray luminosity $L_{{\rm peak}, X} \sim 10^{43} \, \rm erg \, s^{-1}$ given by our model, where $\theta_c=3^\circ$ and $\theta_c=16^\circ$ are obtained from MCMC simulations using multi-wavelength afterglow data of GRB 170817A (see Section~\ref{sec:results}).}
    \label{fig:prompt-peak}
\end{figure*}

We denote by $\mathcal{L}_{i,j}$ the source-frame luminosity per unit solid angle of the $(i,j)$ patch. The corresponding observed luminosity is Doppler-transformed as
\begin{equation}
    \mathcal{L}_{i,j,\rm obs}
    =
    \mathcal{L}_{i,j}\,
    \mathcal{R}_{\mathcal D}^{4}(\theta_{\rm v},\phi_{\rm v},\theta_i,\phi_j).
\end{equation}
The X-ray peak luminosity at different viewing angles is shown in Fig.~\ref{fig:prompt-peak}.

To compute the total emission at a given observer time $t_{\rm obs}$, we sum the contributions from all emitting surface elements lying on the corresponding equal-arrival-time surface. In practice, this is performed by interpolating the emission from the discrete $(\theta_i,\phi_j)$ grid onto a common observer-time grid.
The total observed luminosity per unit solid angle is then
\begin{equation}
    \mathcal{L}_{\rm tot} = \frac{\sum_{i,j} \mathcal{L}_{i,j,\,\rm obs} \, \Delta\Omega_{i,j}}{\sum_{i,j} \Delta\Omega_{i,j}}, 
    \label{eq:L_tot}.
\end{equation}

To obtain isotropic-equivalent quantities in a given energy band, one can integrate the total observed spectrum (Eq.~\ref{eq:N_E_tot}) and light curve (Eq.~\ref{eq:L_tot}), 
\begin{align}
\label{eq:iso_obs}
    E_{\rm iso} &= 4\pi\bar\epsilon = 4\pi\int_{E_1}^{E_2} E\mathcal{N}_{\rm tot}(E) \, dE, \\
    L_{\rm iso} &= 4\pi \mathcal{L}_{\rm tot}.
\end{align}

As the observer moves off-axis from the core of a structured jet, the local emission along the line of sight ($\theta_v$) diminishes significantly relative to the Doppler-shifted contribution from the bright jet core. For viewing angles $\theta_v > \theta_c$, most of the observed emission originates from an intermediate region between the LoS and jet axis at the maximum of $\epsilon \mathcal{R}_\mathcal{D}^3$---effectively the “brightest” visible region (Fig.\ref{fig:doppler}). 
As a result, even when the intrinsic energy distribution is modeled with a Gaussian profile, the observed isotropic-equivalent energy $E_{\rm iso}(\theta_v)$ deviates at large viewing angles $\theta_v \gg \theta_c$ (Fig.~\ref{fig:convergence}).
Furthermore, for observers at large viewing angles ($\theta_v \gg \theta_c$) the local emission is faint, hence the observed spectrum is dominated instead by Doppler-shifted contributions from the bright jet core. 
Consequently, the observed spectrum primarily reflects the properties of the brightest emitting patch---where $E_{\rm iso} \propto \mathcal{R}_\mathcal{D}^3$, $E_p \propto \mathcal{R}_\mathcal{D}$. This leads to a substantially higher observed $E_p$ than predicted by the classical Amati relation, as illustrated in Fig.~\ref{fig:amati}. Notably, this deviation was observed in GRB170817A~\citep{grb170817_fermi,zhang+18_peculiar_grb}.

The model presented here shares conceptual similarities with \cite{ascenzi_20}. Both works describe the curvature (high-latitude) effect of prompt radiation in a structured relativistic outflow and compute the observed light curve by integrating emission over equal-arrival-time surfaces for a generic viewing angle. In this sense, both studies model the same geometric curvature effect.
The main difference lies in the modeling approach. \citet{ascenzi_20} adopt a geometrical prescription for the emission radius, $R_{\rm em}(\theta)=\beta(\theta)ct_{\rm em}$,
while, in our model we define emission surface as $R_{\rm em}(\theta)=\beta(\theta)ct_{\rm em}/(1-\beta(\theta))$.

\subsection{Wind}
\label{sec:wind}
A newly formed magnetar is expected to launch a relativistic, Poynting-flux dominated wind. The bulk Lorentz factor is subject to uncertainties. In the early phase, the hot proto-neutron star may launch a baryon-contaminated neutrino-driven wind. Later, the wind may enter a high-magnetization regime ($\sigma \gg 1$), where baryon loading is minimal. The Lorentz factor of the outflow can quickly approach the wind's termination Lorentz factor ($\Gamma \sim \sigma_0^{1/3}$) and additional acceleration may be possible due to collimation or internal pressure gradient~\citep{granot_11, metzger_11}. 
Nonetheless, further baryon loading is possible, either from the surface or from the ejecta. 
The actual Lorentz factor of the wind is poorly constrained, and we adopt a moderate value based on following considerations: 
\begin{enumerate}
    \item Observationally, the wind emission is manifested as an X-ray plateau, with a lower $E_p$ than prompt emission. Theoretical arguments invoking synchrotron radiation or quasi-thermal emission all require a smaller $\Gamma$ than prompt emission \citep{zhang_02_spectrum, granot_11, beniamini_17},
    \item X-ray flares are believed to be related to late-central engine activities similar to X-ray plateuas. Some constraints on the Lorentz factor plateaus point toward to a moderate value~\citep{yi_15_xray_flare, begue_22, begue_25}, which lends indirect support to the smaller $\Gamma$ for the plateau emission, and
    \item the empirical relation between $\Gamma$ and emission luminosity \citep[e.g.][]{lvj+12} also suggests that the wind luminosity may be moderate.
\end{enumerate}

The X-ray luminosity of the wind is parameterized as
\begin{equation}
\label{eq:L_sd_x}
    L_{X}(t) = \eta_X L_{\rm sd}(t)
\end{equation}
where $\eta_X$ is the efficiency of converting spindown energy to X-ray luminosity and $L_{\rm sd}(t)$ is the spindown luminosity (Eq.~\ref{eq:spindown}). The efficiency factor is poorly constrained and model-dependent, varying from $\eta_X \lesssim 0.2$~\citep{rowlinson+14_efficiency} to $\eta_X > 0.4$~\citep{gao+16_binary_ns} to $\eta_X$ of order $10^{-2}$~\citep{xiao_dai19_efficiency}.
In this work, we adopt $\eta_X=0.1$. Using typical SMNS parameters (Table~\ref{tab:SMNS}) in Eq.~\ref{eq:spindown}, we estimate the X-ray luminosity to be $L_{X}\sim10^{48} \rm \, erg \, s^{-1}$. This corresponds quite well to typical X-ray plateau luminosities and provides a natural explanation for such emission \citep{lu15_ms_magnetar}. For a collapse time $t_c=t_{\rm sd} = 300 \, \rm s$~\citep{ai_20_eos}, 
the total wind energy is $E_X = (3\times10^{50} \rm \, erg)\ \eta_{X, \, -1}^{-1}$. Numerically, we define an isotropic component with $L_{X} = 10^{48} \rm \, erg \, s^{-1}$ and $\Gamma_{\rm wind} = 50$. 

We also note that wind isotropy is not necessarily expected; the accumulation of toroidal magnetic fields generates directional pressure imbalances that redirect initially equatorial flow toward the poles~\citep{begelman_92, konigl_02}.
Numerical simulations have found anisotropic wind evolutions in rapidly-rotating neutron stars~\citep{bucc_06, wang_24} and binary neutron star systems~\citep{skiathas_25}.
Here, we adopt wind isotropy as a simple proof-of-concept model and note that both \texttt{PromptX}~\citep{promptx} (for prompt emission) and \texttt{VegasAfterglow} (for afterglow emission; introduced in the next section) accept any structure of relativistic outflows. 

Unlike in Section~\ref{sec:prompt}, we do not define a spectral model for the X-ray wind. Instead we directly model the on-axis X-ray wind light curve using spindown luminosity (Eq.~\ref{eq:l_sd}) and calculate the observed luminosity per solid angle, $\mathcal{L}_{X, \, i,j, \, \rm obs} = \mathcal{L}_{X,\, i,j} \mathcal{R}_\mathcal{D}^4$, for a given observer at ($\theta_{\rm v}, \phi_{\rm v}$). Then, $\mathcal{L}_{X, \, \rm tot}=\sum_{i,j} \mathcal{L}_{X, \, i,j, \rm \, obs} \Delta\Omega_{i,j} / \sum_{i,j} \Delta\Omega_{i,j}$. Note that as long as ($\theta_{\rm v}, \phi_{\rm v}$) is within the Free zone, the observer is ``on axis'' of the wind and $\mathcal{L}_{X ,\, \rm tot}$ remains nearly constant. Once the observer is beyond the Trapped zone ($\theta > \theta_{\rm cut}$), $\mathcal{R}_\mathcal{D} \neq 1$ and $\mathcal{L}_{X ,\, \rm tot}$ will be attenuated.

\subsection{Afterglow}
\label{sec:afterglow}

\begin{figure}
    \centering
    \includegraphics[width=\linewidth]{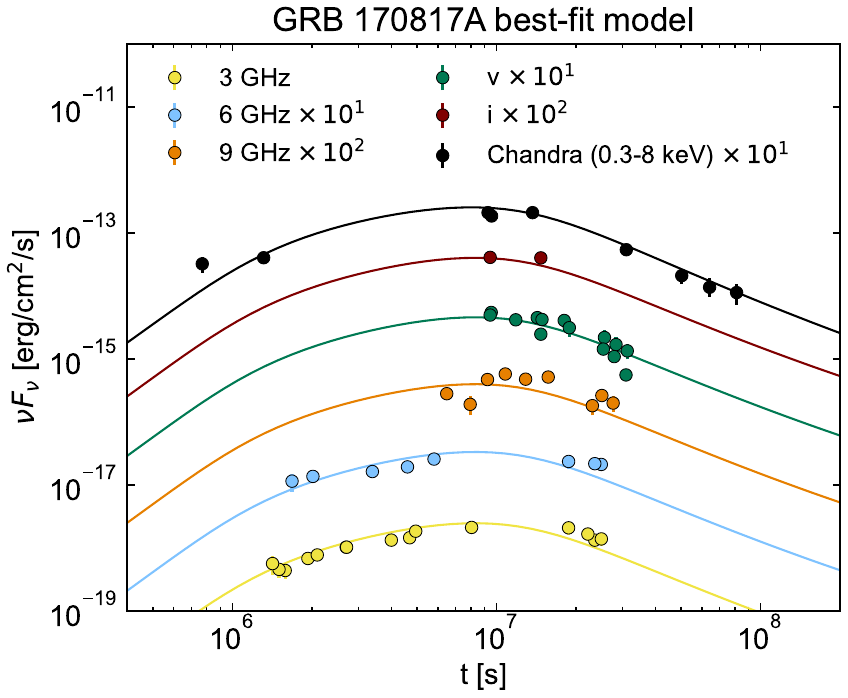}
    \caption{MCMC best-fit afterglow model for a compiled set of GRB 170817A optical~\citep{lyman_18_optical, fong_19_optical, lamb_19_optical}, radio~\citep{hallinan_17, alexander_17, alexander_18, dobie_18_radio_170817, mooley_18_radio}, and X-ray~\citep{troja+17_GW170817_xray, troja_19, troja+20_170817} afterglow data. Model parameters are listed in Table~\ref{tab:model_parameters}.}
    \label{fig:best_fit}
\end{figure}

For both BNS and BH-NS mergers, the GRB jet collides with the circumburst medium and drives a pair of shocks---a relativistic forward shock propagating into the unshocked medium and a reverse shock propagating back into the ejecta~\citep{meszaros_rees_97, mr_99, sari_98, sari+99, zkm_03}. As the jet sweeps up external material, it decelerates, and the shocks accelerate electrons that radiate from radio to gamma-rays, producing the afterglow emission.
The reverse shock is typically short-lived and peaks at early times, while the forward shock persists and dominates at later times.

We use \texttt{VegasAfterglow} \citep{wang_25_vegasafterglow} to model X-ray afterglow light curves for a variety of jet geometries and observers' LoS. 
\texttt{VegasAfterglow} is a high-performance, modular C++ framework wrapped in an intuitive Python interface for GRB afterglow modeling that prioritizes both computational efficiency and physical fidelity. The code self-consistently evolves the dynamics of forward and reverse shocks, supports arbitrary user-defined jet structures and magnetization profiles, accounts for lateral spreading and energy injection, and includes a full treatment of synchrotron and inverse Compton radiation with optional Klein-Nishina corrections. The framework consistently tracks the evolution of GRB jets into the deep Newtonian regime, enabling accurate modeling across the extended afterglow timeline. A full description of the implemented physics and semi-analytic approximations, as well as a comparison with other publicly available afterglow codes can be found in~\cite{wang_25_vegasafterglow}.

For models invoking a magnetar central engine, we account for isotropic energy injection into the jet and the dynamical ejecta using $L_{\rm inj} = (1 - \eta_X)L_{\rm sd}$ with the SMNS model parameters listed in Table~\ref{tab:SMNS}. The injection function operates only within the solid angle subtended by the jet or the ejecta, respectively.

The isotropic, non-relativistic kilonova ejecta would interact with the ambient medium to produce a kilonova afterglow. The emission is weak and usually buried beneath the bright GRB jet afterglow emission. It may, however, emerge when the jet enters the deep Newtonian phase at a much later time. 
An apparent flattening in the late-time X-ray light curve of GRB 170817A was initially interpreted as the possible emergence of a kilonova afterglow \citep{troja+20_170817,hajela_22}.
However, this interpretation was later challenged by \cite{troja_22}, who re-analyzed the data and attributed the observed flattening to an analysis artifact, highlighting the growing tension between jet models and observations.
Recent analysis by \cite{katira_25} supports this reassessment, showing that the X-ray afterglow continues to decline smoothly.
Despite the lack of a confirmed detection in GRB 170817A, we include the kilonova afterglow component in our modeling with \texttt{VegasAfterglow} to quantify its potential contribution and assess the conditions under which it may become observable. We assume an isotropic dynamical ejecta with a total kinetic energy of $\sim 10^{51}$ erg and an average initial velocity of $\sim 0.3c$~\citep{Villar2017,Cowperthwaite2017,Drout2017}.

\section{Results}
\label{sec:results}

\begin{table*}
\centering
\caption{Model parameters for jet prompt and afterglow emission of GRB 170817A obtained from MCMC simulations (see text for details). The value of $E_{\gamma, \rm iso}$ assumes a radiative efficiency of prompt material, $\eta_\gamma = 0.01$, which is consistent with observational constraints~\citep{beniamini_15, beniamini_16_efficiciency, salafia_21_efficiency}}
\label{tab:model_parameters}
\begin{tabular}{|l|ccc|ccccccccc|}
\hline
\hline
\textbf{Model} & \( \theta_c \) [$^\circ$] & \( \theta_v \) [$^\circ$] & \( \theta_{\rm cut} \) [$^\circ$] & \( E_{\rm \gamma,iso} \) [erg] &
\( \Gamma_{\rm 0} \) & \( E_{\rm k,iso} \) [erg] & \( n_{\rm ism} \) [cm$^{-3}$] & \( p \) & \( \varepsilon_e \) & \( \varepsilon_B \) & $\xi_n$ & \(E_{\rm inj} \) \\
\hline
BNS-I, -II & $3$ & $16$ & $20$ & \( 10^{51} \) & 
$200$ & \( 10^{53} \) & $10^{-2.9}$  & $2.13$ & $10^{-2.9}$ & $10^{-2.5}$ & $10^{-1.2}$ & \checkmark \\

BNS-III, -IV & $3$ & $16$ & $20$ & \( 10^{51} \) & 
$200$ & \( 10^{53} \) & $10^{-2.9}$  & $2.13$ & $10^{-2.9}$ & $10^{-2.5}$ & $10^{-1.2}$ & $\boldsymbol{\times}$ \\

BH-NS + TD & $10$ & $16$ & $40$ & \( 10^{50} \) & 
$200$ & \( 10^{52} \) & $10^{-2.9}$  & $2.13$ & $10^{-2.9}$ & $10^{-2.5}$ & $10^{-1.2}$ & $\boldsymbol{\times}$ \\
\hline
\end{tabular}
\end{table*}

The evolution and fate of an NS merger and the associated EM signatures depend on the particular progenitor, post-merger remnant, and observer's LoS to the system. For X-ray prompt and wind emission, we use the numerical model described in Sections~\ref{sec:prompt} and \ref{sec:wind}, and developed in \texttt{PromptX}~\citep{promptx}. For X-ray afterglow emission from the jet and kilonova, we use \texttt{VegasAfterglow}~\citep{wang_25_vegasafterglow}.

GRB 170817A is used as a case study in this work. Using \texttt{VegasAfterglow}, we perform an MCMC simulation for $30,000$ steps and a $30\%$ burn-in on a compiled set of GRB 170817A optical~\citep{lyman_18_optical, fong_19_optical, lamb_19_optical}, radio~\citep{hallinan_17, alexander_17, alexander_18, dobie_18_radio_170817, mooley_18_radio}, and X-ray~\citep{troja+17_GW170817_xray, troja_19, troja+20_170817} afterglow data. 

\begin{figure*}
    \centering
    \includegraphics[width=\linewidth]{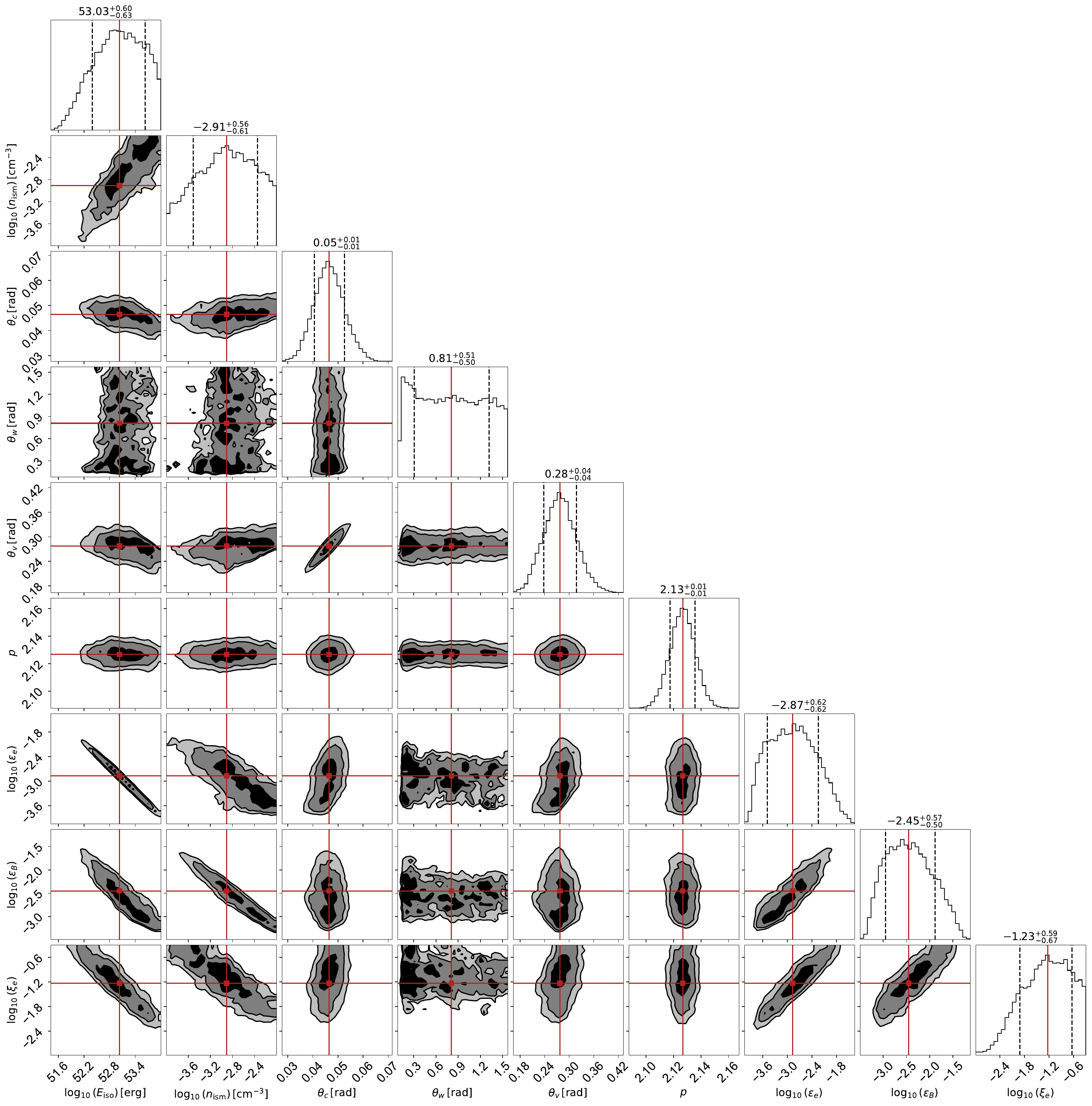}
    \caption{MCMC posterior distribution of afterglow parameters from a compiled set of GRB 170817A optical~\citep{lyman_18_optical, fong_19_optical, lamb_19_optical}, radio~\citep{hallinan_17, alexander_17, alexander_18, dobie_18_radio_170817, mooley_18_radio}, and X-ray~\citep{troja+17_GW170817_xray, troja_19, troja+20_170817} afterglow data}
    \label{fig:mcmc_corner}
\end{figure*}

In our MCMC simulations, the initial Lorentz factor, $\Gamma_0$ is formally unbounded, which can lead to nonphysical values if left unrestricted. Following common practice in the literature, we adopt a fiducial value of $\Gamma_0 = 200$, consistent with other studies of GRB 170817A~\citep{ghirlanda_19_170817, beniamini_20, ryan_24}, to ensure realistic jet dynamics and afterglow modeling. The best-fit model parameters are listed in Table~\ref{tab:model_parameters}, with the corresponding light curves and spectra shown in Fig.~\ref{fig:best_fit}. The MCMC posterior distribution is presented in Fig.~\ref{fig:mcmc_corner}.

In the following subsections, we explore various progenitors, central engines and viewing angles to identify the most plausible scenario of GRB 170817A.

\subsection{BNS Mergers}

We adopt the following convention to label the BNS evolutionary path in order of increasing remnant mass:

\begin{enumerate}
    \item BNS-I---$M_{\rm rem} < M_{\rm TOV}$: The BNS merger will produce a differentially-rotating remnant (i.e., the HMNS ``phase'') that eventually settles into a stable NS that undergoes spindown and survives indefinitely.
    \item BNS-II---$M_{\rm rem} \sim (1.0-1.2)\, M_{\rm TOV}$: The BNS merger remnant will evolve through the HMNS phase and become an SMNS that will ultimately collapse into a BH once rotational energy is sufficiently dissipated by spindown. 
    \item BNS-III---$M_{\rm rem} \sim (1.2-1.3) \, M_{\rm TOV}$: The BNS merger will evolve through the HMNS phase before quickly collapsing into a BH once differential rotation is sufficiently damped by magnetic braking.
    \item BNS-IV---$M_{\rm rem} > 1.3 \, M_{\rm TOV}$: The BNS merger will promptly form a BH.
\end{enumerate}

In all models of BNS mergers, we postulate an opaque ejecta beyond a cutoff angle, $\theta_{\rm cut}$. This parameter, sometimes called $\theta_w$ (wing), is poorly constrained by our MCMC simulations of the afterglow. We adopt $\theta_{\rm cut}= 20^\circ$ based on the ratio $\theta_c:\theta_v:\theta_{\rm cut} \approx 1:6:7$ consistent with afterglow modeling of GRB 170817A by ~\citet{troja_19, troja+20_170817, ryan_24}.

~\cite{wang_24} conducted numerical simulations showing that an initially isotropic wind launched within an ejecta shell interacts with the ejecta and becomes collimated toward the poles. For low ejecta masses ($M_{\rm ej} \sim 10^{-3} M_\odot$), this interaction can widen the polar cavity within seconds. Due to limitations on the simulated duration, they were unable to determine the final opening angle of the funnel. Here we assume that in the steady-state Free zone (after $\sim 10^2 \, \rm s$) the opening angle is $\theta_{\rm cut} = 20^\circ$.

For each evolutionary path, we show the prompt and afterglow X-ray light curves alongside gamma-ray and X-ray observations of GRB 170817A in Figs.~\ref{fig:bns1_lc}--\ref{fig:bns3_lc}, and discuss the implications of the results.

We note that our results are derived solely from EM observations of the prompt and afterglow emissions following a BNS merger. In the case of GW170817/GRB 170817A, constraints on the merger remnant (e.g., maximum mass and collapse time of an SMNS) have been obtained from the ejecta mass and jet-launching delay timescale assuming that only black holes can power short GRBs~\citep{margalit_17, gill_19}. In this paper, we keep open the possibility that SMNSs can power GRB jets, and therefore do not consider these additional constraints.

\subsubsection{BNS-I (BNS $\rightarrow$ HMNS $\rightarrow$ stable NS)}
\label{bns1}

Physically, one expects prompt and afterglow X-ray emission from the GRB jet, as well as extended X-ray emission from the wind of a long-lived remnant. The prompt X-ray emission does not cease immediately after the engine shuts off due to the high-latitude effect \cite{ascenzi_20}, while the extended X-ray emission from the wind decays as $L_{\rm sd} \propto t^{-2}$ (Eq.~\ref{eq:l_sd}).  The kilonova afterglow may be detectable in X-rays at much later times.

\begin{figure*}
    \centering
    \includegraphics[width=0.328\linewidth]{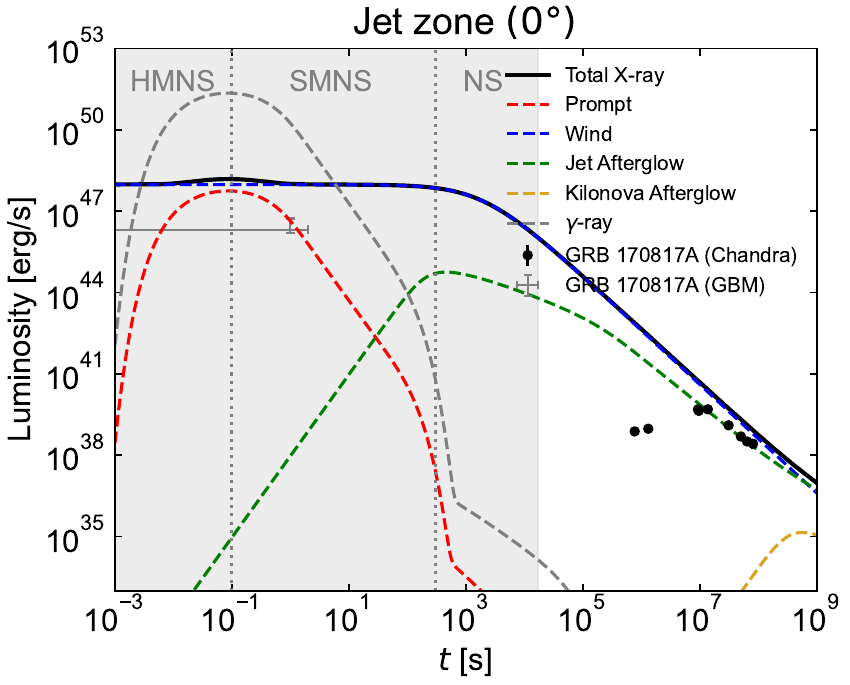}
    \includegraphics[width=0.328\linewidth]{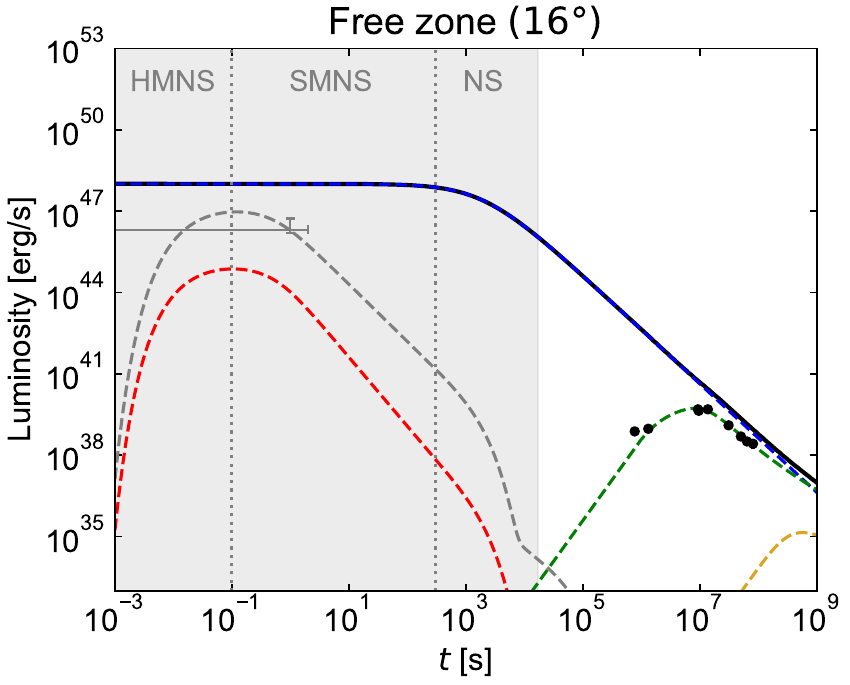}
    \includegraphics[width=0.328\linewidth]{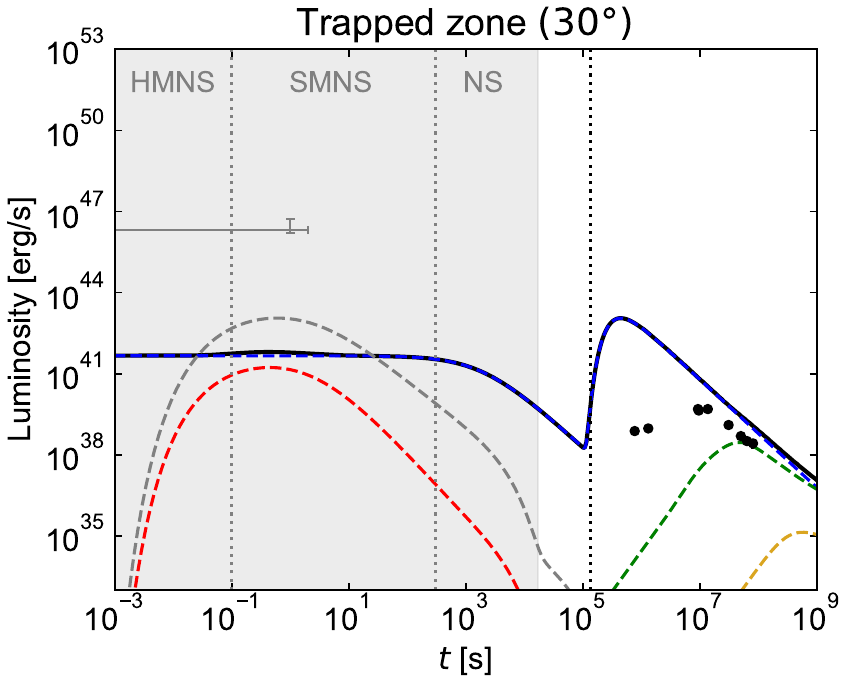}
    \caption{BNS-I model light curves composed of GRB prompt (red), magnetar wind (blue), GRB afterglow (green), and kilonova afterglow (yellow) emissions. Prompt emission is calculated for a Gaussian jet with $\theta_c = 3^\circ$ and $E_{\gamma, \, \rm iso} = 10^{51}$ erg (see Section~\ref{sec:prompt}). Magnetar wind emission is calculated from spindown luminosity (Eq.~\ref{eq:L_sd_x}) and assumed to be isotropic (see Section~\ref{sec:wind}). GRB and kilonova afterglows are calculated using \texttt{VegasAfterglow} (see Section~\ref{sec:afterglow}). We account for an opaque dynamical ejecta by imposing a cutoff, $\theta_{\rm cut} = 20^\circ$, beyond which magnetar wind emission along $\theta_v$ is blocked until the opacity time, $t_{\rm thin}$ (vertical dotted black line). GBM gamma-ray data was retrieved from ~\cite{grb170817_fermi} and \citep{zhang_18}; Chandra X-ray data was retrieved from \cite{,troja+17_GW170817_xray, troja+18, troja_19, troja+20_170817}. The gray shaded region denotes early times without X-ray coverage.
    }
    \label{fig:bns1_lc}
\end{figure*}

Observationally, in X-rays, one would observe various emission signatures depending on the viewing angle,
\begin{itemize}
    \item Jet zone: Bright prompt emission and decaying afterglow from the GRB jet. The extended wind emission becomes observable once the prompt emission sufficiently diminishes. See left panel of Fig.~\ref{fig:bns1_lc}.
    \item Free zone: Faint prompt emission and rising afterglow from the GRB jet. The extended wind emission becomes observable once the prompt emission sufficiently diminishes---in some cases the prompt emission will never exceed the wind emission and only a plateau may be detected. See middle panel of Fig.~\ref{fig:bns1_lc}.
    \item Trapped zone: Extremely faint prompt emission and dim late-time rising afterglow from the GRB jet. The extended wind emission becomes directly observable once the dynamical ejecta becomes optically thin. Doppler-shifted wind emission from the Free zone may be detectable. See right panel of Fig.~\ref{fig:bns1_lc}.
\end{itemize}

The BNS-I scenario can be definitively ruled out in the case of GRB 170817A due to the lack of X-ray emission signatures until the rising jet afterglow at later times~\citep{troja+17_GW170817_xray}.

\subsubsection{BNS-II (BNS $\rightarrow$ HMNS $\rightarrow$ SMNS $\rightarrow$ BH)}
\label{bns2}

Physically, one would expect prompt and afterglow X-rays from the GRB jet and extended X-ray emission from the wind of a long-lived remnant. The extended X-ray emission would cease suddenly at $t_{\rm sd}$ (Eq.~\ref{eq:t_sd}), and only the X-ray afterglow would remain. The kilonova afterglow may be detectable in X-rays at much later times.

\begin{figure*}
    \centering
    \includegraphics[width=0.328\linewidth]{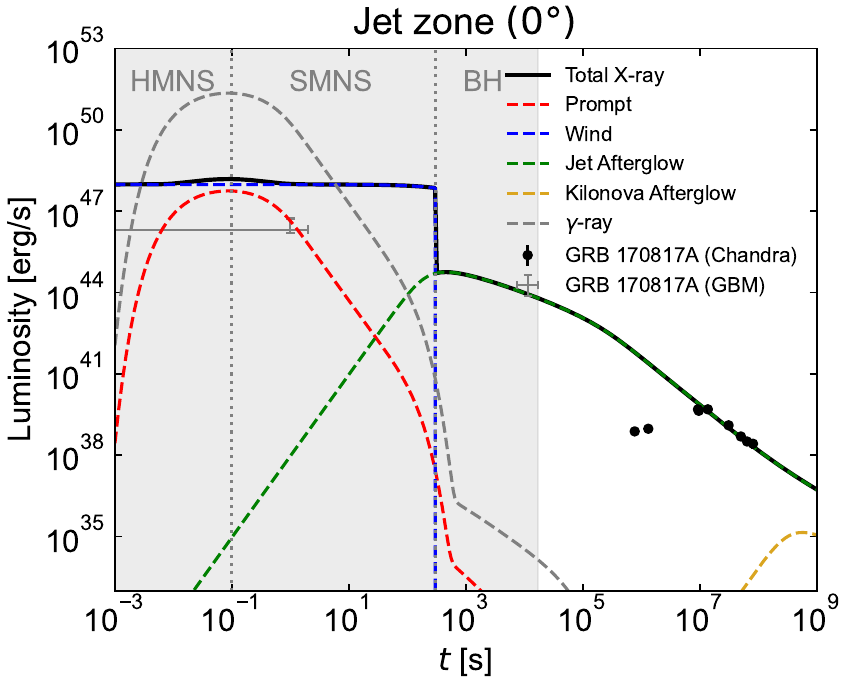}
    \includegraphics[width=0.328\linewidth]{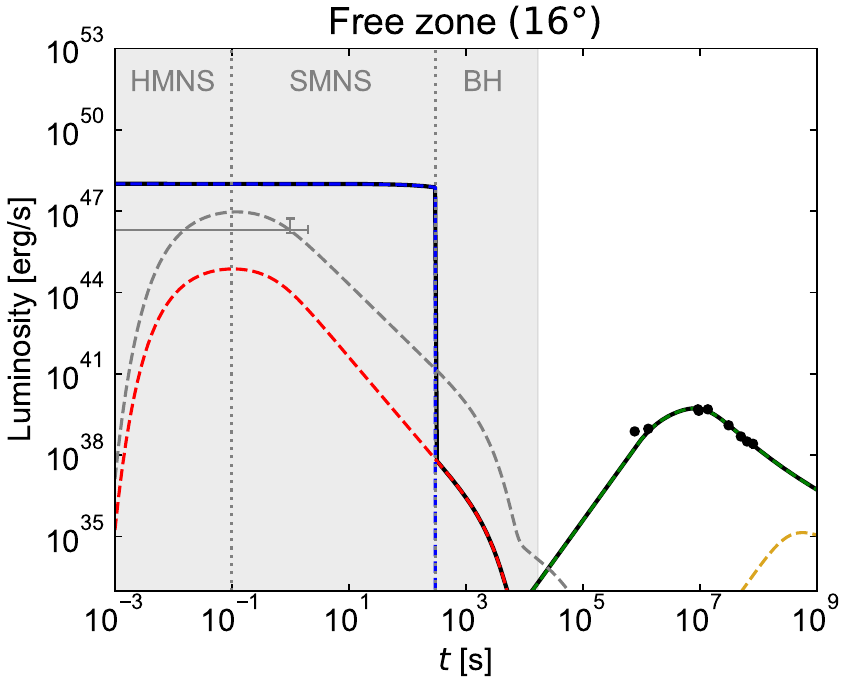}
    \includegraphics[width=0.328\linewidth]{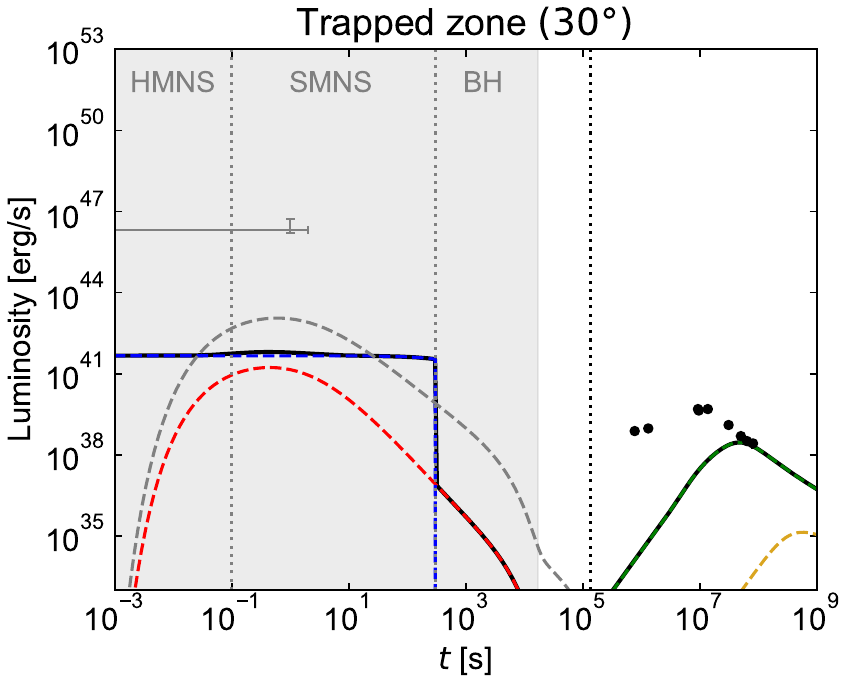}
    \caption{Same as Fig.~\ref{fig:bns1_lc} but for BNS-II.}
    \label{fig:bns2_lc}
\end{figure*}

Observationally, in X-rays, one would observe various emission signatures depending on the viewing angle,
\begin{itemize}
    \item Jet zone: Bright prompt emission and decaying afterglow from the GRB jet. The extended wind emission becomes observable once the prompt emission sufficiently diminishes, and will decay quickly (prescribed by high-latitude emission effects) when the remnant collapses. See left panel of Fig.~\ref{fig:bns2_lc}.
    \item Free zone: Faint prompt emission and rising afterglow from the GRB jet. The extended X-ray wind emission becomes observable once the prompt emission sufficiently diminishes---in some cases the prompt emission will never exceed the wind emission and only a plateau will be observed. The plateau will decay quickly (prescribed by high-latitude emission effects) when the remnant collapses. See middle panel of Fig.~\ref{fig:bns2_lc}.
    \item Trapped zone: Extremely faint prompt emission and dim late-time rising afterglow from the GRB jet. If $t_{\rm thin} > t_c$, the extended wind emission will never be directly observable. See right panel of Fig.~\ref{fig:bns2_lc}.
\end{itemize}

The BNS-II scenario may not be conclusively ruled out for GRB 170817A due to a lack of X-ray coverage at very early times caused by localization delays, large positional uncertainties, and Earth occultation. 
This underscores the importance of continued advances in gravitational-wave localization pipelines and the development of wide-field and rapid-response X-ray detectors to enable prompt follow-up of future BNS mergers.

\subsubsection{BNS-III (BNS $\rightarrow$ HMNS $\rightarrow$ BH)}
\label{bns3}
Physically, one would expect typical prompt and afterglow X-rays from the GRB jet without significant energy injection. The kilonova afterglow may be detectable in X-rays at much later times.

\begin{figure*}
    \centering
    \includegraphics[width=0.328\linewidth]{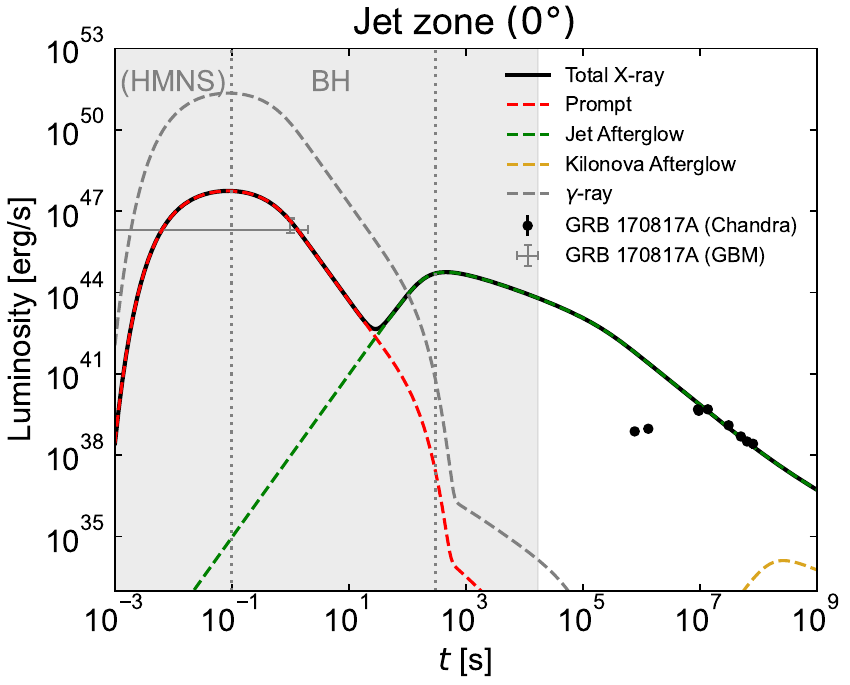}
    \includegraphics[width=0.328\linewidth]{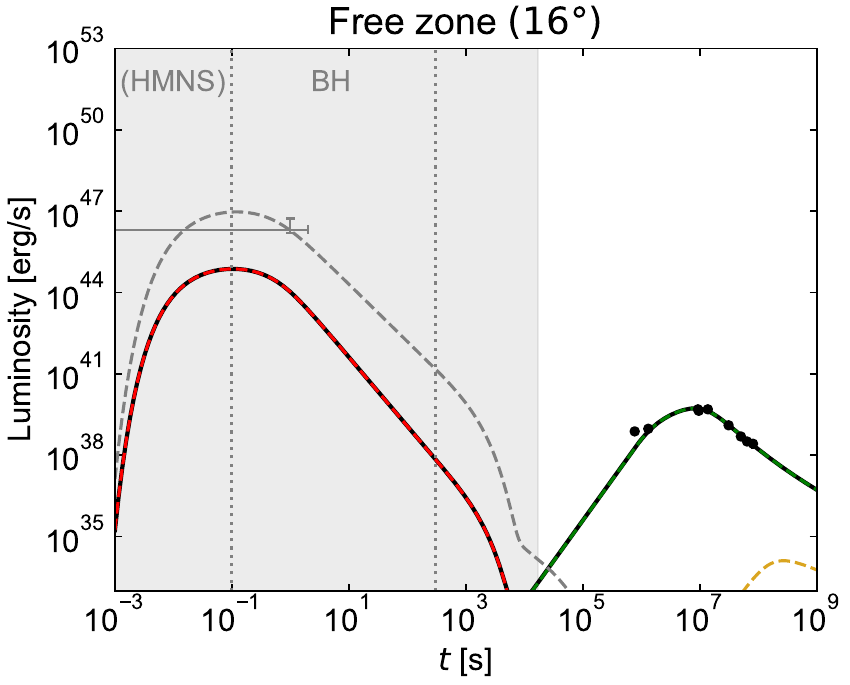}
    \includegraphics[width=0.328\linewidth]{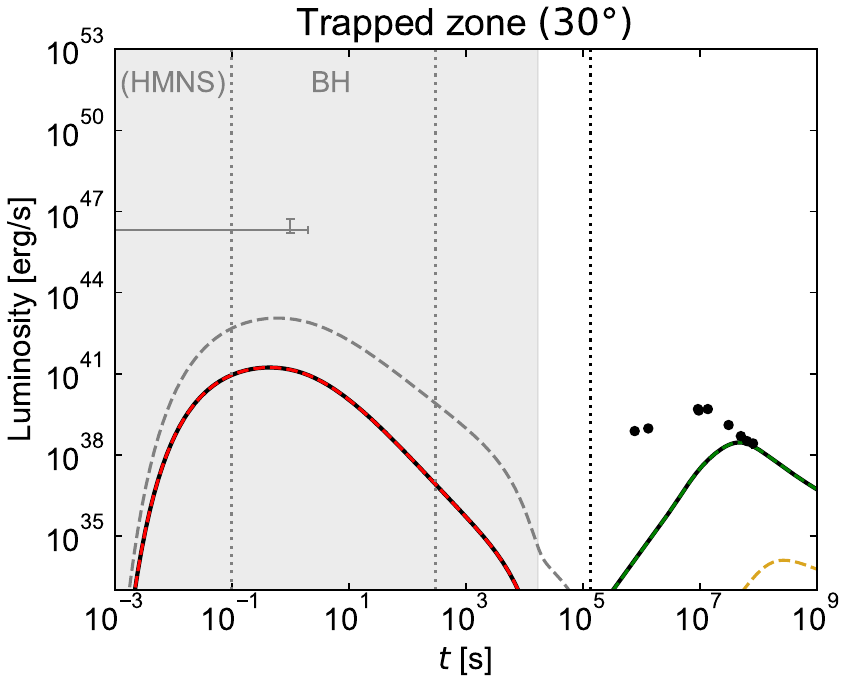}
    \caption{Same as Fig.~\ref{fig:bns1_lc} but for BNS-III and BNS-IV.}
    \label{fig:bns3_lc}
    \end{figure*}

Observationally, in X-rays, one would observe various emission signatures depending on the viewing angle,
\begin{itemize}
    \item Jet zone: Bright prompt emission and decaying afterglow from the GRB jet. See left panel of Fig.~\ref{fig:bns3_lc}.
    \item Free zone: Faint prompt emission and rising afterglow from the GRB jet. See middle panel of Fig.~\ref{fig:bns3_lc}.
    \item Trapped zone: Extremely faint prompt emission and dim late-time rising afterglow from the GRB jet. See right panel of Fig.~\ref{fig:bns3_lc}.
\end{itemize}

The BNS-III scenario can naturally reproduce the light curves of GRB 170817A and does not necessitate a long-lived central engine.

\subsubsection{BNS-IV (BNS $\rightarrow$ BH)}
\label{bns4}
Remnants with masses much larger than the TOV mass limit will be unable to withstand collapse into a BH for an appreciable amount of time.
A jet may still be launched if there is matter outside the event horizon forming an accretion disk, and the associated light curve would likely resemble that of a BNS-III (Fig.~\ref{fig:bns3_lc})

As with the BNS-III scenario, the BNS-IV scenario also naturally reproduces the light curves of GRB 170817A.

\subsection{BH-NS Mergers}
\label{sec:bhns}
X-ray counterparts are only expected from BH-NS mergers that tidally disrupt the NS (BH-NS w/ TD).
Physically, one might expect prompt and afterglow X-rays from a relativistic GRB jet. 
Compared to jets launched from BNS mergers, jets launched from BH-NS mergers w/ TD are likely less collimated~\citep{bromberg_11_propagation, duffell_15_jet, ruiz_18_nsbh_jet, lazzati_perna_19}. Additionally, the dynamical ejecta should be directed primarily in the equatorial direction~\citep{kyutoku_15, kawaguchi_15_bhns, kawaguchi_16_bhns}, which affects the peak flux and time of X-ray signatures from BH-NS mergers~\citep{sadeh_24}. Therefore, we adopt $\theta_c = 10^\circ$ and $\theta_{\rm cut} = 40^\circ$ for a BH-NS w/ TD jet and adjust $E_{\rm iso}$ such that the beaming-corrected energy $E_\gamma = (1 - \cos\theta_c) \, E_{\gamma, \rm iso}$ remains constant. The BH-NS w/ TD jet, while less collimated than its BNS counterpart, has a shallower energy and Lorentz factor profile and larger cutoff angle, thus appearing brighter at large viewing angles.

\begin{figure*}
    \centering
    \includegraphics[width=0.328\linewidth]{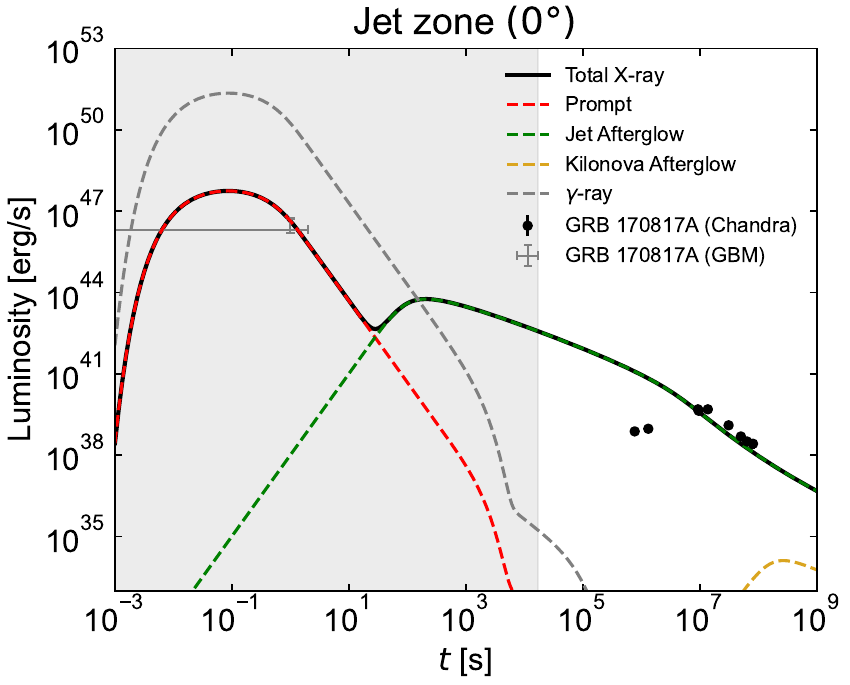}
    \includegraphics[width=0.328\linewidth]{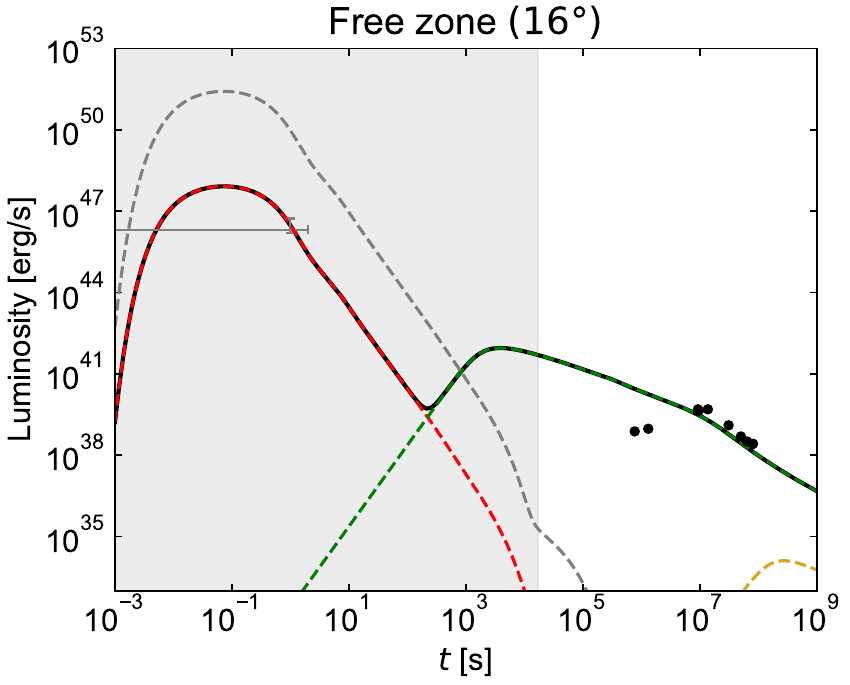}
    \includegraphics[width=0.328\linewidth]{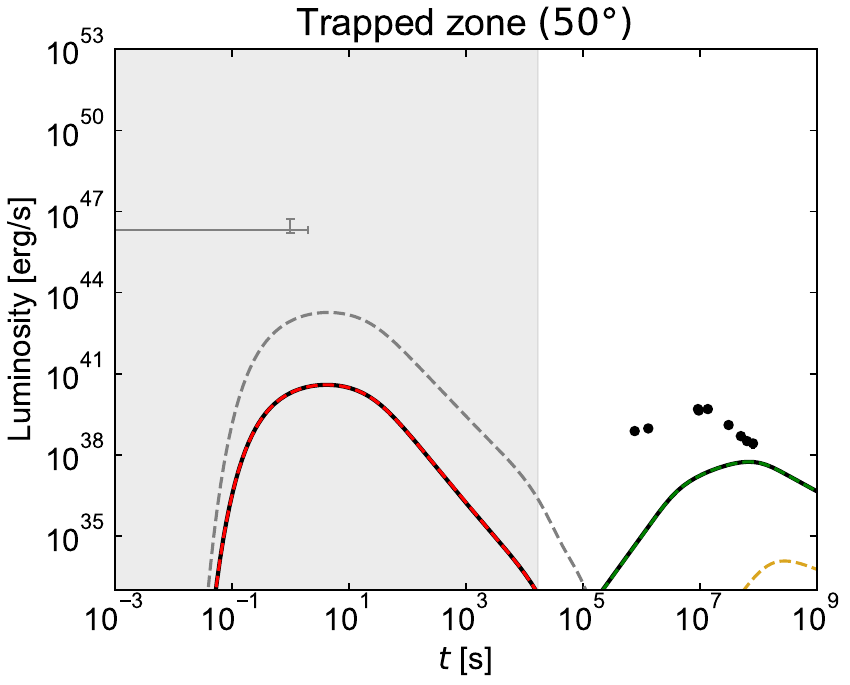}
    \caption{Same as Fig.~\ref{fig:bns1_lc} but for BH-NS w/ TD.}
    \label{fig:bhns_lc}
\end{figure*}

Observationally, in X-rays, one would observe various emission signatures depending on the viewing angle,
\begin{itemize}
    \item Jet zone: Bright prompt emission and decaying afterglow from the GRB jet. Fig.~\ref{fig:bhns_lc} (\textit{left}).
    \item Free zone: Faint prompt emission and rising afterglow from GRB jet. Fig.~\ref{fig:bhns_lc} (\textit{middle}).
    \item Trapped zone: Extremely faint prompt emission and dim late-time rising afterglow from the GRB jet. Fig.~\ref{fig:bhns_lc} (\textit{right}).
\end{itemize}

Due to very strong evidence that GRB 170817A was associated with a BNS merger event---including the GW170817 signal~\citep{GW170817}, associated kilonova~\cite{Cowperthwaite2017}, redshift and localization, and our model parameters for BH-NS mergers---we definitively rule out the BH-NS model for GRB 170817A.

\subsection{The case of GW170817}
The EM counterparts of GW170817, including both prompt and afterglow emission of GRB 170817A, can be consistently interpreted within the framework presented here. However, since the earliest X-ray coverage of the GW170817 final localization region were not until $>10^4 \, \rm s$ trigger, it remains unclear whether this BNS merger was a BNS-II or BNS-III/BNS-IV. Prompt X-ray detections of NS mergers by a rapid-slewing or wide-field X-ray detector can identify the presence of a long-lived central engine and distinguish between these scenarios.

\subsection{X-ray Counterpart Searches to Neutron Star Mergers}
The search for EM counterparts of NS mergers detected by the LIGO-Virgo-KAGRA collaboration (LVK) is critical for advancing understanding of both the merger process and the Type I GRBs. The joint detection of GW170817 and GRB 170817A has opened a new era of multimessenger astronomy, demonstrating the power of combining GW and EM observations.

With the ongoing improvement in sensitivity and detection volume of ground-based GW detectors, including future upgrades and next-generation observatories, we anticipate an increase in the number of detections of NS mergers with associated EM signatures. Therefore, establishing robust and timely follow-up strategies---particularly for the X-ray afterglows---is essential.

For any NS mergers with a potential EM counterpart, X-ray signals may arise from shock interactions, magnetar spin-down, fallback accretion, or relativistic jet afterglows, and they typically evolve on timescales from hundreds of seconds to several years post-merger. Early detection of these signals can provide valuable insight into the nature of the remnant, energy dissipation mechanisms, and the surrounding environment.

Once candidates are localized, rapid follow-up observations using detectors like Swift/XRT~\citep{gehrels_04} have historically proven invaluable for capturing the onset and evolution of X-ray afterglows. Its rapid response capability and sensitivity enable timely monitoring of transient sources. 

The newest generation of wide-field X-ray telescopes, including the Einstein Probe (EP; \cite{yuan_25_ep}) and the Space-based multi-band astronomical Variable Objects Monitor (SVOM; \cite{svom}), boasts rapid response and dedicated sensitivity in the soft X-rays. EP's Wide-field X-ray Telescope (WXT), with its wide field-of-view and high sensitivity in the soft X-ray band ($0.5$--$4$ keV), is designed to promptly detect and localize soft X-ray transients across broad sky regions. Its on-board Follow-up X-ray Telescope (FXT), allows deep follow-up observations in X-rays. SVOM offers complementary multi-wavelength capabilities through its diverse payload: the wide-field ECLAIR ($4-150 \rm \, keV$) enables prompt detections, while the Microchannel X-ray Telescope (MXT; $0.2-10 \rm \, keV$) facilitates rapid X-ray follow-up. Additionally, SVOM is equipped with rapid slewing capability and is integrated with ground-based follow-up networks, enhancing its effectiveness in capturing transient events.

Motivated by these advancing observational capabilities and the need to prepare for future detections, here we present the intrinsic X-ray peak luminosity and time of prompt and afterglow emissions from NS mergers across all viewing angles. The results can be used jointly to provide a complete picture of X-ray emission signatures across wide observational epochs.

Fig.~\ref{fig:prompt-peak} shows the peak X-ray luminosity of prompt emission from NS mergers with EM counterparts observed at different viewing angles, for several jet opening angles and assuming a constant beaming-corrected energy $E_{\gamma, \rm tot}$. For BNS mergers, the jet opening angle is typically narrow, with $\theta_c \sim 3^\circ$ for GRB 170817A, whereas for BH-NS mergers, the opening angle is likely larger, $\theta_c\sim 10^\circ$~\citep{bromberg_11_propagation, duffell_15_jet, ruiz_18_nsbh_jet, lazzati_perna_19}. 
The true jet opening angle derived for GRB170817A remains unclear, with differing results from various methods~\citep{gottlieb_18, mooley_18, troja_19, ghirlanda_19_170817, ryan_24}. This work is generally consistent with models invoking a single-component narrow GRB jet.

\begin{figure*}
    \centering
    \includegraphics[width=\linewidth]{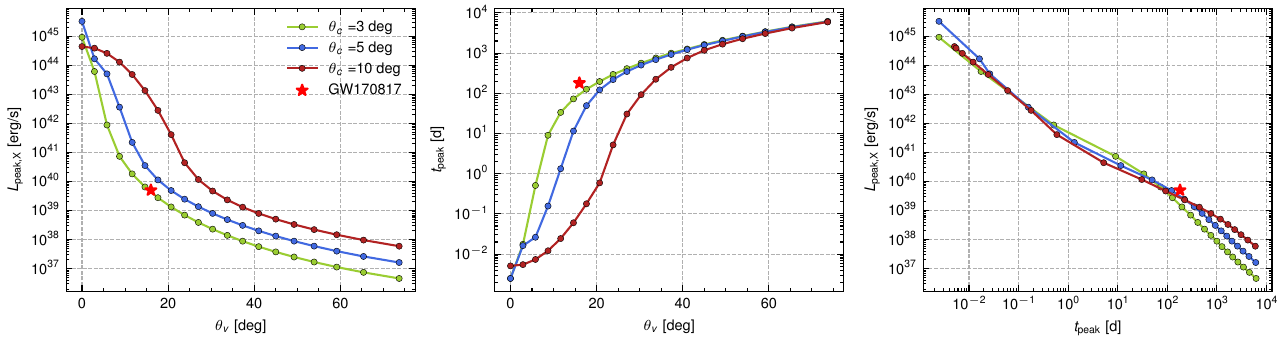}
    \caption{X-ray afterglow peak luminosity (\textit{left panel}) and peak time (\textit{middle panel}) as functions of viewing angle for several jet opening angles and assuming a constant beaming-corrected energy $E_{\gamma, \rm tot}$. The \textit{right panel} shows the correlation between peak luminosity and peak time. The beaming-corrected energy is fixed at $10^{51}$ erg. The red star represents the best-fit values from our MCMC simulations for the X-ray afterglow peak luminosity and time for GRB 170817A. X-ray data was retrieved from \cite{troja+20_170817}.}
    \label{fig:afterglow-peak}
\end{figure*}

\begin{figure*}
    \centering
    \includegraphics[width=\linewidth]{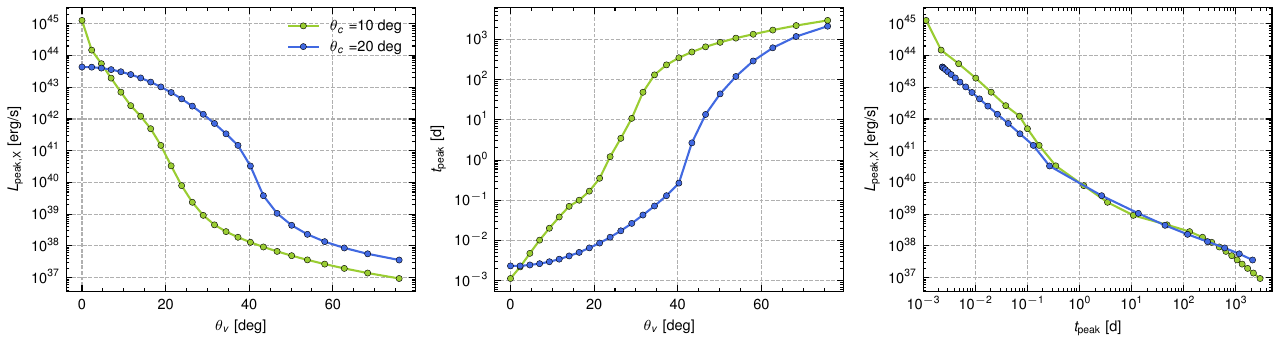}
    \caption{Same as Fig.~\ref{fig:afterglow-peak}, but for BH-NS w/ TD.}
    \label{fig:afterglow-peak-bhns}
\end{figure*}

\begin{figure}
    \centering
    \includegraphics[width=\linewidth]{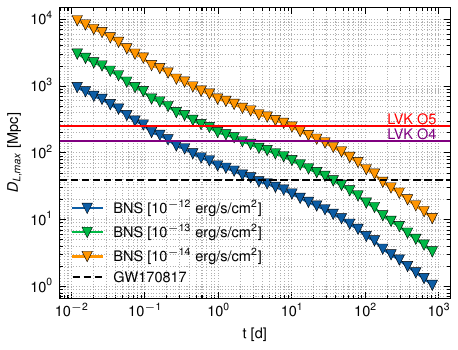}
    \caption{Detection horizon for X-ray (0.3-10 keV) afterglows from BNS mergers as a function of time, for given detector sensitivities. The model parameters are listed in Tab.~\ref{tab:model_parameters}.}
    \label{fig:Dl_c}
\end{figure}

With constraints from prompt X-ray emission, one may estimate the X-ray afterglow peak time and luminosity. Fig.~\ref{fig:afterglow-peak} shows the X-ray afterglow peak luminosity (\textit{left panel}) and peak time (\textit{middle panel}) as functions of the viewing angle $\theta_v$ for several jet opening angles and assuming a constant beaming-corrected energy $E_{\gamma, \rm tot}$. 
With the inferred luminosity distance from LVK, observers may estimate the detection horizon for X-ray counterparts of NS mergers based on their detector's sensitivity and time since the GW trigger (Fig.~\ref{fig:Dl_c}).

Here we summarize the basic X-ray counterpart search workflow,
\begin{itemize}
    \item \textbf{Detection}: An NS merger event is identified by either a GW signal detected by a GW observatory or a Type I GRB detected by a gamma-ray detector.
    
    \item \textbf{Prompt X-ray Analysis}: Data from X-ray detectors that were covering the localization region during the trigger time should be examined for coincident events.  If no events are found, upper limits can help constrain the prompt X-ray emission.
    
    \item \textbf{Geometric Constraints}: Constraints on $\theta_c$, $\theta_v$, and $\theta_{\rm cut}$ for a GRB jet with beaming-corrected energy $E_{\gamma, \rm tot}$ corresponding to GRB 170817A can be derived from X-ray detections and upper limits using the model described in Section~\ref{sec:prompt} and illustrated in Fig.~\ref{fig:prompt-peak}.
    
    \item \textbf{X-ray Afterglow Predictions}: Given a fixed set of model parameters and geometric constraints from prompt emission, the X-ray afterglow characteristics (e.g., peak time and flux) can be estimated (Fig.~\ref{fig:afterglow-peak}).\footnote{We emphasize that these predictions are highly sensitive to the chosen physical parameters and are only representative of bursts with properties similar to those assumed in the model. In this work, we adopt parameters representative of GRB 170817A, so the resulting predictions are specific to GW170817/GRB 170817A-like events.}
    
    \item \textbf{Coordinated Follow-up}: X-ray follow-up observations can be planned based on the predicted afterglow properties. Using Fig.~\ref{fig:Dl_c}, one can estimate the optimal time window for detecting an X-ray counterpart of an NS merger with known luminosity distance, given one's detector sensitivity. It suggests that prompt follow-up observations are very important. In case of non-detection at a certain epoch, one can estimate the luminosity distance horizon beyond which continued follow-up may no longer yield detections. 
\end{itemize}

Additional detections of NS mergers with X-ray counterparts will be a major success of multimessenger astronomy. 

Fig.~\ref{fig:Dl_c} shows the detection horizon of a BNS-merger GRB afterglow with luminosity comparable to GRB 170817A. 
Here, we focus on the afterglow as it sets the long-term window for coordinated follow-up observations and, through its properties, provides constraints on the prompt emission and central engine activity that cannot be obtained from short-lived signals alone.

At any given epoch, there exists a range of viewing angles for which the afterglow peak emission remains above a given detector's flux threshold. The detection horizon curves therefore represent the largest distance at which an off-axis afterglow exactly peaks at the threshold at a given time. Equivalently, for a fixed source distance, the afterglow would remain potentially detectable until the horizon curve drops below that distance; this includes cases where the off-axis afterglow is still rising and its peak has not yet occurred.
For example, if EP-FXT (with sensitivity of about $10^{-14} \ {\rm erg/s/cm^2}$) does not detect an X-ray counterpart of a BNS merger like GW170817/GRB 170817A within one day, then follow-up observations may no longer yield detections if the source is further than $\sim 300$ Mpc. Equivalently, for a source with luminosity distance $D_L \sim 140 \rm \, Mpc$ (e.g., obtained from a GW trigger as in GW170817) EP-FXT could successfully detect an X-ray counterpart within $\sim 100$ days.

The expected distribution of intrinsic events across viewing zones (jet, free, and trapped) is proportional to the solid angle subtended by each zone, $\Omega = 4\pi\int_{\theta_1}^{\theta_2} \sin\theta \, d\theta = 4\pi(\cos\theta_2 - \cos\theta_1)$, because observers are expected to be distributed isotropically. Therefore, the fractions of events observed in each zone can be expressed as
\begin{align}
f_{\rm jet} &= \frac{\Omega_{\rm jet}}{4\pi}=1 - \cos \theta_{\rm c} \\
f_{\rm free} &= \frac{\Omega_{\rm free}}{4\pi}=\cos \theta_{\rm c} - \cos \theta_{\rm cut} \\
f_{\rm trapped} &= \frac{\Omega_{\rm trapped}}{4\pi}=\cos \theta_{\rm cut}.
\end{align}
For BNS mergers ($\theta_c = 3^\circ$ and $\theta_{\rm cut} = 20^\circ$), $f_{\rm jet} = 0.0014$, $f_{\rm free} = 0.0589$, and $f_{\rm trapped} = 0.9397$. This yields an intrinsic jet : free : trapped zone ratio of $\approx1:42:671$.
For BHNS mergers ($\theta_c = 10^\circ$ and $\theta_{\rm cut} = 40^\circ$), $f_{\rm jet} = 0.0152$, $f_{\rm free} = 0.2188$, and $f_{\rm trapped} = 0.7660$. This yields an intrinsic jet:free:trapped zone ratio of $\approx 1:14:50$. We note, however, the observed jet:free:trapped zone ratios could be much different from these intrinsic values. This is because given a fixed detector sensitivity, the chance of detecting faint emission in the trapped or even free zones is much lower than that of detecting emission from the jet zone. A full analysis of detectability effects is beyond the scope of this paper. 

\section{Conclusion}
\label{sec:conclusion}
In this paper, we have systematically investigated the X-ray signatures associated with NS mergers, considering various progenitors and central engines. We have introduced a method for modeling prompt X-ray light curves and spectra for any jet profile and observer viewing angle. 

Our analysis specifically explored four distinct BNS merger scenarios categorized by remnant mass and post-merger evolutionary pathways: stable neutron star formation (BNS-I), delayed black hole collapse following a supramassive neutron star phase (BNS-II), rapid black hole formation post-hypermassive neutron star phase (BNS-III), and prompt black hole formation (BNS-IV). Each scenario revealed characteristic X-ray emission patterns: notably, long-lived magnetar remnants (BNS-I and II) produce distinct, prolonged X-ray plateau phases powered by magnetar spin-down emissions, which significantly differentiate them from scenarios leading directly or rapidly to black hole formation.

Furthermore, we highlighted how viewing geometry dramatically influences observed X-ray signatures, classifying observation zones into jet-aligned (brightest emission), free (moderately dimmed emission), and trapped (strongly dimmed or delayed emission due to ejecta opacity) zones. Such geometrical considerations are useful for interpreting X-ray data from multimessenger observations, especially in conjunction with gravitational wave detections.

We performed a case study of GRB 170817A using MCMC simulations on multi-wavelength data, and showed that the prompt and afterglow frameworks developed in this paper are self-consistent and in agreement with similar models in the literature.

Last, we proposed a basic workflow for identifying and characterizing X-ray counterparts of GW-detected NS mergers, emphasizing rapid and coordinated follow-up strategies utilizing X-ray observatories such as Einstein Probe, SVOM, and Swift. This proactive observational approach will enhance our ability to detect, characterize, and interpret the physical processes underlying NS mergers.

\section*{Code Availability}
The code used to compute prompt X-ray emission from a relativistic jet and a magnetar-powered wind (Sections~\ref{sec:prompt} and~\ref{sec:wind}) is publicly available under the MIT license on GitHub (\href{https://github.com/cjules0609/PromptX}{github.com/cjules0609/PromptX}) and archived in Zenodo~\citep{promptx}.
The code for modeling GRB afterglow emission (Section~\ref{sec:afterglow}) is publicly available under the BSD 3-clause license on Github: \href{https://github.com/YihanWangAstro/VegasAfterglow}{github.com/YihanWangAstro/VegasAfterglow}~\citep{wang_25_vegasafterglow}.

Both repositories provide the complete source code, accompanying documentation, and example scripts, enabling other researchers to reproduce the results and apply these methods to their own analyses.

\section*{Acknowledgements}
The authors acknowledge NASA 80NSSC23M0104, NASA 80NSSC20M0043, the Nevada Center for Astrophysics, and a UNLV Top-Tier Doctoral Graduate Research Assistantship (TTDGRA) for support.

\bibliography{xray_grb}{}
\bibliographystyle{aasjournal}

\end{document}